\documentclass[english,prb,aps,preprintnumbers,twocolumn]{revtex4}
\usepackage[T1]{fontenc}
\usepackage[latin9]{inputenc}
\usepackage{array}
\usepackage{units}
\usepackage{amsmath}
\usepackage{amssymb}
\usepackage{graphicx}
\usepackage{esint}
\PassOptionsToPackage{normalem}{ulem}
\usepackage{ulem}

\makeatletter

\providecommand{\tabularnewline}{\\}

\@ifundefined{textcolor}{}
{%
 \definecolor{BLACK}{gray}{0}
 \definecolor{WHITE}{gray}{1}
 \definecolor{RED}{rgb}{1,0,0}
 \definecolor{GREEN}{rgb}{0,1,0}
 \definecolor{BLUE}{rgb}{0,0,1}
 \definecolor{CYAN}{cmyk}{1,0,0,0}
 \definecolor{MAGENTA}{cmyk}{0,1,0,0}
 \definecolor{YELLOW}{cmyk}{0,0,1,0}
}

\makeatother

\usepackage{babel}
\begin{document}

\title{Steady state conductance in a double quantum dot array: The nonequilibrium
equation-of-motion Green function approach }

\author{Tal J. Levy and Eran Rabani}

\affiliation{School of Chemistry, The Sackler Faculty of Exact Sciences, Tel Aviv
University, Tel Aviv 69978, Israel}
\begin{abstract}
We study steady state transport through a double quantum dot array
using the equation-of-motion approach to the nonequilibrium Green
functions formalism. This popular technique relies on uncontrolled
approximations to obtain a closure for a hierarchy of equations, however
its accuracy is questioned. We focus on $4$ different closures, $2$
of which were previously proposed in the context of the single quantum
dot system (Anderson impurity model) and were extended to the double
quantum dot array, and develop $2$ new closures. Results for the
differential conductance are compared to those attained by a master
equation approach known to be accurate for weak system-leads couplings
and high temperatures. While all $4$ closures provide an accurate
description of the Coulomb blockade and other transport properties
in the single quantum dot case, they differ in the case of the double
quantum dot array, where only one of the developed closures provides
satisfactory results. This is rationalized by comparing the poles
of the Green functions to the exact many-particle energy differences
for the isolate system. Our analysis provides means to extend the
equation-of-motion technique to more elaborate models of large bridge
systems with strong electronic interactions.
\end{abstract}
\maketitle

\section{Introduction\label{sec:Introduction}}

The interest in transport through conjugated molecules has grown in
recent years in light of their potential applications in electronic
and optoelectronic devices.~\citep{Kelley2004,Gur2005} While certain
transport properties can be treated within a non-interacting picture
via the tight binding approximation combined with the Landauer formalism,~\citep{Landauer1957}
often the description of transport requires the inclusion of many-body
electron-electron and/or electron-phonon correlations.~\citep{Haug1996}
For very simple and small systems, introducing such correlations can
be done, for example, by means of time-dependent numerical renormalization-group
techniques,~\citep{white_density_1992,schmitteckert_nonequilibrium_2004}
many-body wavefunction approaches,~\citep{Thoss2011} diagrammatic
techniques to real time path integral formulation,~\citep{muehlbacher_real-time_2008,weiss_iterative_2008,eckel_comparative_2010,Segal10}
or reduced dynamic methods.~\citep{leijnse_kinetic_2008,Cohen2011}
The treatment of correlations becomes a greater theoretical challenge
in systems with many electronic states driven away from equilibrium,~\citep{Matthew2000,Leeuwen2009GW}
where the computational cost of numerical techniques increases rapidly
beyond current capabilities. 

A central framework dealing with transport in large systems is the
nonequilibrium Green functions (NEGF) formalism.~\citep{Schwinger1961a,Keldysh1964,Datta2000,Xue2002}
The equation-of-motion (EOM) method is one of the more basic ways
to calculate the Green functions (GF) of an interacting quantum system.
Its main advantages are the simplicity and relatively mild scaling
with the number of electrons, under simple truncation schemes.~\citep{Pals1996}
The EOM nonequilibrium Green function formalism provides a qualitative
description of transport phenomena in strongly correlated systems,
such as the Coulomb blockade effect~\citep{Lacroix1981,Meir1991,Song2007}
and the Kondo effect in quantum dots.~\citep{Meir1993,Galperin2007a}
However, questions regarding the validity of the EOM approach have
been raised.~\citep{Kashcheyevs2006,Levy2013} For example, it has
been shown to violate the Friedel sum rule~\citep{Langreth66} near
the Kondo regime~\citep{Kashcheyevs2006} and basic Green function
symmetry relations away from the Kondo regime.~\citep{Levy2013}
In the latter case, symmetry relations can be restored and at least
for the Anderson impurity model,\citep{Anderson1961} the approach
recovers the Kondo peaks and provides a quantitative description of
resonant transport.~\citep{Levy2013}

In this paper we study the role of different approximate closures
to the EOM of the NEGF formalism on steady state properties (namely,
the differential conductance) for a double quantum dot (QD) array,
coupled to two macroscopic leads (an extended Hubbard model,~\citep{Hubbard1963,Vermeulen1995}
also known as the double Anderson model~\citep{Jayaprakash81}).
Four closures are examined; two already proposed~\citep{Song2007,Joshi2000}
(approximation $1$ and $4$ described in subsection~\ref{sub:Case1}
and~\ref{sub:Case4}, respectively) and two developed in this work
(approximation $2$ and $3$ described in subsections~\ref{sub:Case2}
and~\ref{sub:Case3}, respectively). The results obtained from the
different closures were compared to the results attained using a many-particle
Master Equation (ME) approach~\citep{Oppenheim1977} adequate for
weak hybridization (system-leads couplings) and high temperatures.~\citep{Datta1994,Datta1995}
In contrast to the simplest case of a single site model (Anderson
impurity model) in which different closures beyond the simplest Hartree
approximation scheme~\citep{Lacroix1981} yield very similar transport
results~\citep{Pals1996} (steady state current and differential
conductance as a function of the applied bias voltage) at high temperatures,
we show that this is not the case for the double QD array, where different
closures yield very different steady state currents and differential
conductance curves. 

The performance of the different closures is analyzed in terms of
the poles of the GFs in comparison to the exact many-body result for
the isolated system. We find that one of the closures developed in
this work provides the most accurate description of the poles and
also the best overall agreement with the ME approach for all parameters
studied in this work. While these results are encouraging, a word
of caution is in place. It is clear that the conclusions drawn from
the performance of the different closures for the single site model
cannot be extended directly to the two site model, in analogy, a suitable
closure for the two site model may fail in the larger systems. Thus,
the study of larger arrays of QDs will require analysis along the
lines sketched in this work. 

The paper is organized as follows: in Sec.~\ref{sec:Theory} we present
the double QD model Hamiltonian, provide a short description of the
equation-of-motion technique and a detailed description of the different
approximate closures to the EOM. Sequential labeling of the different
approximate closures refers to the order of the closure. Results and
discussion are given in Sec.~\ref{sec:Results} for the cases of
the symmetric and asymmetric bridges. In Sec.~\ref{sec:Concluding-remarks}
we conclude.

\section{Theory\label{sec:Theory}}

\subsection{Model Hamiltonian\label{sub:Model-Hamiltonian}}

\begin{figure}[t]
\centering{}\includegraphics[width=8.5cm]{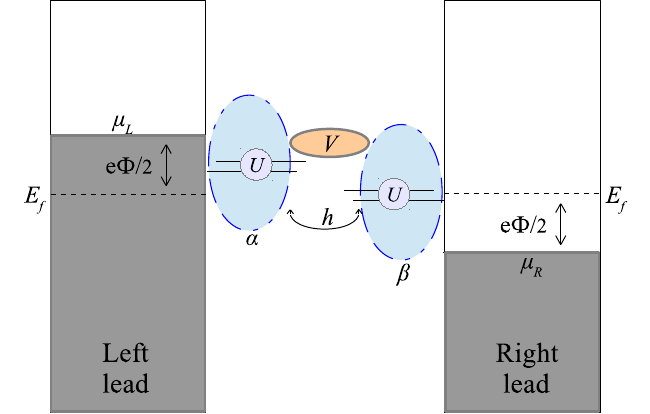}\caption{\label{fig:symmetricBridgeSketch}A sketch of the double QD bridge.
See main text for the definition of all quantities.}
\end{figure}

We consider a system of two coupled QDs array connected to two macroscopic
leads, as sketched in figure~\ref{fig:symmetricBridgeSketch}. The
Hamiltonian has the following general form:
\begin{equation}
\hat{H}=\hat{H}_{B}+\hat{H}_{S}+\hat{H}_{i},
\end{equation}
with $\hat{H}_{B}$ describing the macroscopic leads (left and right
contacts), $\hat{H}_{S}$ describes the system of interest, and $\hat{H}_{i}$
is the interaction Hamiltonian between the system and the leads. The
leads (left ($L$) and right ($R$)) are modeled as infinite non-interacting
fermionic baths,~\citep{Meirav1990,Altshuler1991,Wees1998} and are
assumed to be each at its own equilibrium, characterized by chemical
potentials $\mu_{L}$ and $\mu_{R}$, where the difference $\mu_{L}-\mu_{R}=e\Phi$
is the applied voltage bias. The leads' Hamiltonian is given by:
\begin{equation}
\hat{H}_{B}=\sum_{\sigma,k\in\{L,R\}}\varepsilon_{k\sigma}c_{k\sigma}^{\dagger}c_{k\sigma},
\end{equation}
where $\epsilon_{k\sigma}$ is the energy of a free electron in the
left or right lead, in momentum state $k$ and spin $\sigma$. The
operators $c_{k\sigma}/c_{k\sigma}^{\dagger}$ are the annihilation/creation
operators of such an electron. The double QD system is described by
an extended Hubbard model:~\citep{Hubbard1963,Jayaprakash81}
\begin{eqnarray}
\hat{H}_{S} & = & \sum_{\sigma,m\in\left\{ \alpha,\beta\right\} }\varepsilon_{m\sigma}n_{m\sigma}+U\sum_{m}n_{m\uparrow}n_{m\downarrow}\nonumber \\
 &  & +V\sum_{\sigma,\sigma'}n_{\alpha\sigma}n_{\beta\sigma'}+h\sum_{\sigma}\left(d_{\alpha\sigma}^{\dagger}d_{\beta\sigma}+h.c.\right),
\end{eqnarray}
where $n_{\alpha\sigma}=d_{\alpha\sigma}^{\dagger}d_{\alpha\sigma}$
is the number operator of the electron occupying site (dot) $\alpha$
with spin $\sigma$ and energy $\varepsilon_{\alpha\sigma}$, $U$
is the repulsion energy between two electrons on the same site with
opposite spins (intra-dot repulsion), $V$ is the repulsion energy
between two electrons on different sites (inter-dot repulsion), and
$h$ is the coupling strength for electron hopping between the two
sites. The interaction between the system and the contacts is simply
given by the tunneling Hamiltonian:~\citep{Caroli1971} 
\begin{equation}
\hat{H}_{i}=\sum_{\sigma,k\in L}\left(t_{k\alpha}^{\sigma}c_{k\sigma}^{\dagger}d_{\alpha\sigma}+h.c.\right)+\sum_{\sigma,k\in R}\left(t_{k\beta}^{\sigma}c_{k\sigma}^{\dagger}d_{\beta\sigma}+h.c.\right).
\end{equation}
The parameter $t_{km}^{\sigma}$ represents the coupling strength
(hybridization) between the system and the leads, and the index $m$
runs over the site index $\left\{ \alpha,\beta\right\} $.

\subsection{Equation of motion\label{sub:EOM}}

The above model consists of an interacting system coupled to two electron
reservoirs with specified chemical potentials and temperatures. In
order to obtain a solution to this many-body out of equilibrium problem,
we resort to the EOM approach within the NEGF formalism.~\citep{Haug1996,jauho-2005,Galperin2007a}
We begin by defining the contour ordered GF, $\hat{G}_{\alpha\beta}\left(t_{2},t_{1}\right)=-\frac{i}{\hbar}\left\langle T_{C}\hat{\Psi}_{\alpha}\left(t_{2}\right)\hat{\Psi}_{\beta}^{\dagger}\left(t_{1}\right)\right\rangle $,
where $\hat{\Psi}/\hat{\Psi}^{\dagger}$ are the system's annihilation/creation
field operators, and $T_{C}$ is the contour time ordering operator.~\citep{Haug1996,Leeuwen2009}
The EOM for the contour ordered GF~\citep{Niu1999a,Marques2006}
is obtained from the Heisenberg EOM for a Heisenberg operator $\frac{\mbox{d}}{\mbox{d}t}\hat{A}\left(t\right)=\frac{i}{\hbar}\left[\hat{H}\left(t\right),\hat{A}_{H}\left(t\right)\right]+\frac{\partial}{\partial t}\hat{A}_{H}\left(t\right),$
where $\left[\hat{A},\hat{B}\right]=\hat{A}\hat{B}-\hat{B}\hat{A}$.
A full description of the system requires the knowledge of the retarded,
advanced and lesser (distribution) GFs:~\citep{Mahan1990}
\begin{eqnarray}
\hat{G}_{\alpha\beta}^{r}\left(t_{2},t_{1}\right) & = & -\frac{i}{\hbar}\theta\left(t_{2}-t_{1}\right)\left\langle \left\{ \hat{\Psi}_{\alpha}\left(t_{2}\right),\hat{\Psi}_{\beta}^{\dagger}\left(t_{1}\right)\right\} \right\rangle ,\nonumber \\
\hat{G}_{\alpha\beta}^{a}\left(t_{2},t_{1}\right) & = & \frac{i}{\hbar}\theta\left(t_{1}-t_{2}\right)\left\langle \left\{ \hat{\Psi}_{\alpha}\left(t_{2}\right),\hat{\Psi}_{\beta}^{\dagger}\left(t_{1}\right)\right\} \right\rangle ,\nonumber \\
\hat{G}_{\alpha\beta}^{<}\left(t_{2},t_{1}\right) & = & \frac{i}{\hbar}\left\langle \hat{\Psi}_{\beta}^{\dagger}\left(t_{1}\right)\hat{\Psi}_{\alpha}\left(t_{2}\right)\right\rangle ,\label{eq:GF definitions}
\end{eqnarray}
where $\left\{ A,B\right\} $ is the anti-commutator of $A$ and $B$.
These real time GFs can be extracted from the contour ordered GF using
the Langreth rules.~\citep{Langreth76,Haug1996} Except for very
simple models (e.g. see Refs.~\onlinecite{Swenson2011} and~\onlinecite{Wilner2012}),
the EOMs of the NEGF will produce ``new'' and higher order GFs that
need to be evaluated. In general this leads (after a few iterations)
to a non tractable hierarchy of equations. To obtain a working closure
one has to truncate the resulting set of equations and/or decouple
the higher order GFs and express them via lower order ones. Recently
it has been shown that such a procedure may lead to symmetries violations
that the GFs must obey by definition~\citep{Levy2013} and a routine
to restore back the symmetries was suggested.~\citep{Levy2013} In
what follows we examine $4$ different closures to the EOMs of the
NEGF for the Hamiltonian discussed in subsection~\ref{sub:Model-Hamiltonian},
and apply the symmetry restoration scheme to circumvent the inherent
flaw of the EOMs approach. Our starting point is the contour ordered
GF:
\begin{equation}
\hat{G}_{\alpha\beta}^{\sigma\sigma}\left(t,t'\right)=-\frac{i}{\hbar}\left\langle T_{C}d_{\alpha\sigma}\left(t\right)d_{\beta\sigma}^{\dagger}\left(t'\right)\right\rangle .
\end{equation}
Writing down its EOM and expressing the result in Fourier space (as
we are interested in steady state properties) we find that the single
particle GFs obey: 
\begin{eqnarray}
G_{\alpha\beta}^{\sigma\sigma}\left(\omega\right) & = & \left(\hbar\omega-\varepsilon_{\alpha\sigma}-\Sigma_{\alpha\sigma}^{0}\left(\omega\right)\right)^{-1}\nonumber \\
 &  & \times\left(\delta_{\alpha\beta}+hG_{\beta\beta}^{\sigma\sigma}\left(\omega\right)+V\mathbb{G}_{\beta\alpha\beta}^{\bar{\sigma}\sigma\sigma}\left(\omega\right)\right.\nonumber \\
 &  & \,\,\,\,\,\,\,\,\left.+V\mathbb{G}_{\beta\alpha\beta}^{\sigma\sigma\sigma}\left(\omega\right)+U\mathbb{G}_{\alpha\alpha\beta}^{\bar{\sigma}\sigma\sigma}\left(\omega\right)\right),
\end{eqnarray}
where $\Sigma_{\alpha\sigma}^{0}\left(\omega\right)=\sum_{k}\left|t_{k\alpha}^{\sigma}\right|^{2}\left(\hbar\omega-\varepsilon_{k\sigma}\right)^{-1}$
is the tunneling self energy (the label ``$0$'' refers to the limits
$U,V\rightarrow0$ where the total self-energy is given by $\Sigma_{\alpha\sigma}^{0}\left(\omega\right)$)
and $\mathbb{G}_{\alpha\beta\gamma}^{\tau\sigma\sigma}\left(\omega\right)$
is the Fourier transform of the $2$-particle GF $\hat{\mathbb{G}}_{\alpha\beta\gamma}^{\tau\sigma\sigma}\left(t,t'\right)=-\frac{i}{\hbar}\left\langle T_{C}n_{\alpha\tau}\left(t\right)d_{\beta\sigma}\left(t\right)d_{\gamma\sigma}^{\dagger}\left(t'\right)\right\rangle $.
Calculating the EOMs for the higher order ($2$-particle) GFs will
give rise to other $2$-particle and $3$-particle GFs:
\begin{eqnarray}
\mathbb{G}_{\alpha\beta\gamma}^{\bar{\sigma}\sigma\sigma}\left(\omega\right) & = & \left(\hbar\omega-\varepsilon_{\beta\sigma}-V-\Sigma_{\beta\sigma}^{0}\left(\omega\right)\right)^{-1}\nonumber \\
 &  & \times\left(\delta_{\gamma\beta}\left\langle n_{\alpha,\bar{\sigma}}\right\rangle +h\mathbb{G}_{\alpha\beta\beta\gamma}^{\bar{\sigma}\bar{\sigma}\sigma\sigma}\left(\omega\right)\right.\nonumber \\
 &  & -h\mathbb{G}_{\beta\alpha\beta\gamma}^{\bar{\sigma}\bar{\sigma}\sigma\sigma}\left(\omega\right)+h\mathbb{G}_{\alpha\alpha\gamma}^{\bar{\sigma}\sigma\sigma}\left(\omega\right)\nonumber \\
 &  & +\sum_{k}\left(t_{k\alpha}^{\bar{\sigma}}\mathbb{F}_{\alpha k\beta\gamma}^{\bar{\sigma}\bar{\sigma}\sigma\sigma}\left(\omega\right)-t_{k\alpha}^{\bar{\sigma}*}\mathbb{F}_{k\alpha\beta\gamma}^{\bar{\sigma}\bar{\sigma}\sigma\sigma}\left(\omega\right)\right)\nonumber \\
 &  & +\sum_{k}t_{k\beta}^{\sigma}\mathbb{F}_{\alpha k\gamma}^{\bar{\sigma}\sigma\sigma}\left(\omega\right)+U\mathbf{G}_{\alpha\beta\beta\gamma}^{\bar{\sigma}\bar{\sigma}\sigma\sigma}\left(\omega\right)\nonumber \\
 &  & \left.+V\mathbf{G}_{\alpha\alpha\beta\gamma}^{\bar{\sigma}\sigma\sigma\sigma}\left(\omega\right)\right),
\end{eqnarray}
\begin{eqnarray}
\mathbb{G}_{\alpha\alpha\gamma}^{\bar{\sigma}\sigma\sigma}\left(\omega\right) & = & \left(\hbar\omega-\varepsilon_{\alpha\sigma}-U-\Sigma_{\alpha\sigma}^{0}\left(\omega\right)\right)^{-1}\nonumber \\
 &  & \times\left(\delta_{\gamma\alpha}\left\langle n_{\alpha,\bar{\sigma}}\right\rangle +h\mathbb{G}_{\alpha\beta\alpha\gamma}^{\bar{\sigma}\bar{\sigma}\sigma\sigma}\left(\omega\right)\right.\nonumber \\
 &  & -h\mathbb{G}_{\beta\alpha\alpha\gamma}^{\bar{\sigma}\bar{\sigma}\sigma\sigma}\left(\omega\right)+h\mathbb{G}_{\alpha\beta\gamma}^{\bar{\sigma}\sigma\sigma}\left(\omega\right)\nonumber \\
 &  & +\sum_{k}\left(t_{k\alpha}^{\bar{\sigma}}\mathbb{F}_{\alpha k\alpha\gamma}^{\bar{\sigma}\bar{\sigma}\sigma\sigma}\left(\omega\right)-t_{k\alpha}^{\bar{\sigma}*}\mathbb{F}_{k\alpha\alpha\gamma}^{\bar{\sigma}\bar{\sigma}\sigma\sigma}\left(\omega\right)\right)\nonumber \\
 &  & +\sum_{k}t_{k\alpha}^{\sigma}\mathbb{F}_{\alpha k\gamma}^{\bar{\sigma}\sigma\sigma}\left(\omega\right)+V\mathbf{G}_{\alpha\beta\alpha\gamma}^{\bar{\sigma}\bar{\sigma}\sigma\sigma}\left(\omega\right)\nonumber \\
 &  & \left.+V\mathbf{G}_{\alpha\beta\alpha\gamma}^{\bar{\sigma}\sigma\sigma\sigma}\left(\omega\right)\right),
\end{eqnarray}
and 
\begin{eqnarray}
\mathbb{G}_{\alpha\beta\gamma}^{\sigma\sigma\sigma}\left(\omega\right) & = & \left(\hbar\omega-\varepsilon_{\beta\sigma}-V-\Sigma_{\beta\sigma}^{0}\left(\omega\right)\right)^{-1}\nonumber \\
 &  & \times\left(\left(\delta_{\beta\gamma}\left\langle n_{\alpha,\sigma}\right\rangle -\delta_{\alpha\gamma}\left\langle d_{\alpha\sigma}^{\dagger}d_{\beta,\sigma}\right\rangle \right)\right.\nonumber \\
 &  & +h\mathbb{G}_{\beta\alpha\gamma}^{\sigma\sigma\sigma}\left(\omega\right)+U\mathbf{G}_{\alpha\beta\beta\gamma}^{\sigma\bar{\sigma}\sigma\sigma}\left(\omega\right)\nonumber \\
 &  & +\sum_{k}\left(t_{k\sigma}^{\sigma}\mathbb{F}_{\alpha k\beta\gamma}^{\sigma\sigma\sigma\sigma}\left(\omega\right)-t_{k\sigma}^{\sigma*}\mathbb{F}_{k\alpha\beta\gamma}^{\sigma\sigma\sigma\sigma}\left(\omega\right)\right)\nonumber \\
 &  & \left.+\sum_{k}t_{k\gamma}^{\sigma}\mathbb{F}_{\alpha k\gamma}^{\sigma\sigma\sigma}\left(\omega\right)+V\mathbf{G}_{\alpha\alpha\beta\gamma}^{\sigma\bar{\sigma}\sigma\sigma}\left(\omega\right)\right).
\end{eqnarray}
In the above equations, $\mathbb{G}_{\alpha\beta\gamma\delta}^{\bar{\sigma}\bar{\sigma}\sigma\sigma}\left(\omega\right)$,
$\mathbf{G}_{\alpha\beta\gamma\delta}^{\tau s\sigma\sigma}\left(\omega\right)$,
$\mathbb{F}_{\alpha k\gamma}^{\tau\sigma\sigma}\left(\omega\right)$,
$\mathbb{F}_{\alpha k\beta\gamma}^{\tau\tau\sigma\sigma}\left(\omega\right)$
and $\mathbb{F}_{k\alpha\beta\gamma}^{\tau\tau\sigma\sigma}\left(\omega\right)$
are the Fourier transforms of $\hat{\mathbb{G}}_{\alpha\beta\gamma\delta}^{\bar{\sigma}\bar{\sigma}\sigma\sigma}\left(t,t'\right)=-\frac{i}{\hbar}\left\langle T_{C}d_{\alpha\bar{\sigma}}^{\dagger}\left(t\right)d_{\beta\bar{\sigma}}\left(t\right)d_{\gamma\sigma}\left(t\right)d_{\delta\sigma}^{\dagger}\left(t'\right)\right\rangle $,
$\mathbf{\hat{G}}_{\alpha\beta\gamma\delta}^{\tau s\sigma\sigma}\left(t,t'\right)=-\frac{i}{\hbar}\left\langle T_{C}n_{\alpha\tau}\left(t\right)n_{\beta s}\left(t\right)d_{\gamma\sigma}\left(t\right)d_{\delta\sigma}^{\dagger}\left(t'\right)\right\rangle $,
$\mathbb{\hat{F}}_{\alpha k\gamma}^{\tau\sigma\sigma}\left(t,t'\right)=-\frac{i}{\hbar}\left\langle T_{C}n_{\alpha\tau}\left(t\right)c_{k\sigma}\left(t\right)d_{\gamma\sigma}^{\dagger}\left(t'\right)\right\rangle $,
$\mathbb{\hat{F}}_{\alpha k\beta\gamma}^{\tau\tau\sigma\sigma}\left(t,t'\right)=-\frac{i}{\hbar}\left\langle T_{C}d_{\alpha\tau}^{\dagger}\left(t\right)c_{k\tau}\left(t\right)d_{\beta\sigma}\left(t\right)d_{\gamma\sigma}^{\dagger}\left(t'\right)\right\rangle $
and $\mathbb{\hat{F}}_{k\alpha\beta\gamma}^{\tau\tau\sigma\sigma}\left(t,t'\right)=-\frac{i}{\hbar}\left\langle T_{C}c_{k\tau}^{\dagger}\left(t\right)d_{\alpha\tau}\left(t\right)d_{\beta\sigma}\left(t\right)d_{\gamma\sigma}^{\dagger}\left(t'\right)\right\rangle $
respectively. To continue, one has to formulate the equations for
these new GFs, which in turn will lead to other (higher order) GFs.
This infinite hierarchy of equations needs to be truncated at a certain
level, a process which is referred to as ``closure''. In general,
closures cannot be improved systematically. Furthermore, it is often
difficult to assess, a priori, the accuracy of a given closure. We
now discuss several different closures which are physically motivated,
tractable, and some are commonly used in the context of transport.

\subsubsection{Approximation 1\label{sub:Case1}}

Following the derivation given in Ref.~\onlinecite{Song2007}, the
following approximations are made: (a) all $3$-particle GFs are set
to zero, (b) simultaneous tunneling of electrons of opposite spins
are neglected, (c) GFs mixing leads and system operators are decoupled
so~\citep{Bulka2004} $\hat{\mathbb{F}}\left(t,t'\right)=-\frac{i}{\hbar}\left\langle T_{C}c_{k\sigma}\left(t\right)n_{\alpha\tau}\left(t\right)d_{\beta\sigma}^{\dagger}\left(t'\right)\right\rangle \approx-\frac{i}{\hbar}t_{k\gamma}^{\sigma}\int\mbox{d}t_{1}\hat{g}_{k}\left(t,t_{1}\right)\left\langle T_{C}d_{\gamma\sigma}\left(t_{1}\right)n_{\alpha\tau}\left(t_{1}\right)d_{\beta\sigma}^{\dagger}\left(t'\right)\right\rangle $
where $\left(i\hbar\frac{\partial}{\partial t}-\varepsilon_{k\sigma}\right)\hat{g}_{k}\left(t,t_{1}\right)=\delta\left(t-t_{1}\right)$,
which is obtained by assuming that $n_{\alpha\tau}\left(t\right)$
is constant (as is the case in steady state). (d) The remaining $2$-particle
GFs of the form $\hat{\mathbb{G}}_{\alpha\beta\gamma}^{\tau\sigma\sigma}\left(t,t'\right)$
are decoupled so $\hat{\mathbb{G}}_{\alpha\beta\gamma}^{\tau\sigma\sigma}\left(t,t'\right)=\left\langle n_{\alpha\tau}\left(t\right)\right\rangle \hat{G}_{\beta\gamma}^{\sigma\sigma}\left(t,t'\right)$.
Assumption (c) is equivalent to treating the coupling to the leads
up to the second order with respect to $t_{km}^{\sigma}$. It neglects
processes necessary to qualitatively capture the Kondo effect,~\citep{Cronenwett1998,Swirkowicz2003,Bulka2004}
yet results are predicted to be reliable for temperatures above the
Kondo temperature ($T_{K}$).~\citep{Meir1993,Schiller1995} The
resulting equations are given by (for brevity, we omit the implicit
dependence on $\omega$):

\begin{eqnarray}
G_{\alpha\alpha}^{\sigma\sigma} & = & \left(\hbar\omega-\varepsilon_{\alpha\sigma}-\Sigma_{\alpha\sigma}^{0}\right)^{-1}\nonumber \\
 &  & \times\left(1+hG_{\beta\alpha}^{\sigma\sigma}+U\mathbb{G}_{\alpha\alpha\alpha}^{\bar{\sigma}\sigma\sigma}+V\mathbb{G}_{\beta\alpha\alpha}^{\sigma\sigma\sigma}+V\mathbb{G}_{\beta\alpha\alpha}^{\bar{\sigma}\sigma\sigma}\right),\nonumber \\
G_{\beta\alpha}^{\sigma\sigma} & = & \left(\hbar\omega-\varepsilon_{\beta\sigma}-\Sigma_{\beta\sigma}^{0}\right)^{-1}\nonumber \\
 &  & \times\left(hG_{\alpha\alpha}^{\sigma\sigma}+U\mathbb{G}_{\beta\beta\alpha}^{\bar{\sigma}\sigma\sigma}+V\mathbb{G}_{\alpha\beta\alpha}^{\sigma\sigma\sigma}+V\mathbb{G}_{\alpha\beta\alpha}^{\bar{\sigma}\sigma\sigma}\right),\nonumber \\
\label{eq:1-p GF case1}
\end{eqnarray}
\begin{eqnarray}
\left(\hbar\omega-\varepsilon_{\beta\sigma}-U-\Sigma_{\beta\sigma}^{0}\right)\mathbb{G}_{\beta\beta\alpha}^{\bar{\sigma}\sigma\sigma} & = & h\left\langle n_{\beta\bar{\sigma}}\right\rangle G_{\alpha\alpha}^{\sigma\sigma},\nonumber \\
\left(\hbar\omega-\varepsilon_{\beta\sigma}-V-\Sigma_{\beta\sigma}^{0}\right)\mathbb{G}_{\alpha\beta\alpha}^{\sigma\sigma\sigma} & = & h\left\langle n_{\beta\sigma}\right\rangle G_{\alpha\alpha}^{\sigma\sigma}-\left\langle d_{\alpha\sigma}^{\dagger}d_{\beta\sigma}\right\rangle ,\nonumber \\
\left(\hbar\omega-\varepsilon_{\beta\sigma}-V-\Sigma_{\beta\sigma}^{0}\right)\mathbb{G}_{\alpha\beta\alpha}^{\bar{\sigma}\sigma\sigma} & = & h\left\langle n_{\alpha\bar{\sigma}}\right\rangle G_{\alpha\alpha}^{\sigma\sigma},\nonumber \\
\left(\hbar\omega-\varepsilon_{\alpha\sigma}-U-\Sigma_{\alpha\sigma}^{0}\right)\mathbb{G}_{\alpha\alpha\alpha}^{\bar{\sigma}\sigma\sigma} & = & \left\langle n_{\alpha\bar{\sigma}}\right\rangle +h\left\langle n_{\alpha\bar{\sigma}}\right\rangle G_{\beta\alpha}^{\sigma\sigma},\nonumber \\
\left(\hbar\omega-\varepsilon_{\alpha\sigma}-V-\Sigma_{\alpha\sigma}^{0}\right)\mathbb{G}_{\beta\alpha\alpha}^{\sigma\sigma\sigma} & = & \left\langle n_{\beta\sigma}\right\rangle +h\left\langle n_{\alpha\sigma}\right\rangle G_{\beta\alpha}^{\sigma\sigma},\nonumber \\
\left(\hbar\omega-\varepsilon_{\alpha\sigma}-V-\Sigma_{\alpha\sigma}^{0}\right)\mathbb{G}_{\beta\alpha\alpha}^{\bar{\sigma}\sigma\sigma} & = & \left\langle n_{\beta\bar{\sigma}}\right\rangle +h\left\langle n_{\beta\bar{\sigma}}\right\rangle G_{\beta\alpha}^{\sigma\sigma}.\nonumber \\
\label{eq:2-p GF case1}
\end{eqnarray}
In general the GFs depend on the expectation values of $\left\langle n_{\gamma\tau}\right\rangle =\left\langle d_{\gamma\tau}^{\dagger}d_{\gamma\tau}\right\rangle $
and $\left\langle d_{\alpha\sigma}^{\dagger}d_{\beta\sigma}\right\rangle $,
which are determined by means of the lesser GF: 
\begin{equation}
\left\langle d_{\alpha\sigma}^{\dagger}d_{\beta\sigma}\right\rangle =-\frac{i\hbar}{2\pi}\int_{-\infty}^{\infty}\left(G_{\beta\alpha}^{\sigma\sigma}\left(\omega\right)\right)^{<}\mbox{d}\omega,
\end{equation}
thus, a self consistent calculation is required.

\subsubsection{Approximation 2\label{sub:Case2}}

A seemingly better approximation scheme is one that relaxes the last
mean-field approximation (assumption ``d'') described in subsection~\ref{sub:Case1}
and the $2$-particle GFs of the form $\hat{\mathbb{G}}_{\alpha\beta\gamma}^{\tau\sigma\sigma}\left(t,t'\right)=-\frac{i}{\hbar}\left\langle T_{C}n_{\alpha\tau}\left(t\right)d_{\beta\sigma}\left(t\right)d_{\gamma\sigma}^{\dagger}\left(t'\right)\right\rangle $
are not decoupled but treated fully. Thus, while equation (\ref{eq:1-p GF case1})
remains the same, the equations for the $2$-particle GFs will now
be given by:
\begin{eqnarray}
\left(\hbar\omega-\varepsilon_{\beta\sigma}-V-\Sigma_{\beta\sigma}^{0}\right)\mathbb{G}_{\alpha\beta\alpha}^{\bar{\sigma}\sigma\sigma} & = & h\mathbb{G}_{\alpha\alpha\alpha}^{\bar{\sigma}\sigma\sigma},\nonumber \\
\left(\hbar\omega-\varepsilon_{\beta\sigma}-V-\Sigma_{\beta\sigma}^{0}\right)\mathbb{G}_{\alpha\beta\alpha}^{\sigma\sigma\sigma} & = & h\mathbb{G}_{\beta\alpha\alpha}^{\sigma\sigma\sigma}-\left\langle d_{\alpha\sigma}^{\dagger}d_{\beta,\sigma}\right\rangle ,\nonumber \\
\left(\hbar\omega-\varepsilon_{\beta\sigma}-U-\Sigma_{\beta\sigma}^{0}\right)\mathbb{G}_{\beta\beta\alpha}^{\bar{\sigma}\sigma\sigma} & = & h\mathbb{G}_{\beta\alpha\alpha}^{\bar{\sigma}\sigma\sigma},\nonumber \\
\left(\hbar\omega-\varepsilon_{\alpha\sigma}-U-\Sigma_{\alpha\sigma}^{0}\right)\mathbb{G}_{\alpha\alpha\alpha}^{\bar{\sigma}\sigma\sigma} & = & \left\langle n_{\alpha\bar{\sigma}}\right\rangle +h\mathbb{G}_{\alpha\beta\alpha}^{\bar{\sigma}\sigma\sigma},\nonumber \\
\left(\hbar\omega-\varepsilon_{\alpha\sigma}-V-\Sigma_{\alpha\sigma}^{0}\right)\mathbb{G}_{\beta\alpha\alpha}^{\sigma\sigma\sigma} & = & \left\langle n_{\beta\sigma}\right\rangle +h\mathbb{G}_{\alpha\beta\alpha}^{\sigma\sigma\sigma},\nonumber \\
\left(\hbar\omega-\varepsilon_{\alpha\sigma}-V-\Sigma_{\alpha\sigma}^{0}\right)\mathbb{G}_{\beta\alpha\alpha}^{\bar{\sigma}\sigma\sigma} & = & \left\langle n_{\beta\bar{\sigma}}\right\rangle +h\mathbb{G}_{\beta\beta\alpha}^{\bar{\sigma}\sigma\sigma}.\label{eq:2-p GF case2}
\end{eqnarray}
As noted above, the retarded, advanced and lesser GFs can now be evaluated
using Langreth theorem~\citep{Langreth76} and the expectation values
($\left\langle n_{\gamma\tau}\right\rangle =\left\langle d_{\gamma\tau}^{\dagger}d_{\gamma\tau}\right\rangle $
and $\left\langle d_{\alpha\sigma}^{\dagger}d_{\beta\sigma}\right\rangle $)
are determined via the lesser GF.

\subsubsection{Approximation 3\label{sub:Case3}}

A more complete treatment of the $2^{\mbox{nd}}$ order GFs requires
relaxing assumption ``b'' in addition to assumption ``d'' (see
previous subsection), described in subsection~\ref{sub:Case1}. Namely,
we attend to the $2$-particle GFs that describe simultaneous tunneling
of electrons of opposite spins in the double QD system, $\hat{\mathbb{G}}_{\alpha\beta\alpha\beta}^{\bar{\sigma}\bar{\sigma}\sigma\sigma}\left(t,t'\right)=-\frac{i}{\hbar}\left\langle T_{C}d_{\alpha\bar{\sigma}}^{\dagger}\left(t\right)d_{\beta\bar{\sigma}}\left(t\right)d_{\alpha\sigma}\left(t\right)d_{\beta\sigma}^{\dagger}\left(t'\right)\right\rangle $.
Again, the only changes are in the equations for the $2$-particle
GFs, and the resulting equations are given by:
\begin{eqnarray}
\mathbb{G}_{\alpha\beta\alpha}^{\bar{\sigma}\sigma\sigma} & = & \left(\hbar\omega-\varepsilon_{\beta\sigma}-V-\Sigma_{\beta\sigma}^{0}\right)^{-1}\nonumber \\
 &  & \times\left(h\mathbb{G}_{\alpha\alpha\alpha}^{\bar{\sigma}\sigma\sigma}+h\mathbb{G}_{\alpha\beta\beta\alpha}^{\bar{\sigma}\bar{\sigma}\sigma\sigma}-h\mathbb{G}_{\beta\alpha\beta\alpha}^{\bar{\sigma}\bar{\sigma}\sigma\sigma}\right),\nonumber \\
\mathbb{G}_{\alpha\beta\alpha}^{\sigma\sigma\sigma} & = & \left(\hbar\omega-\varepsilon_{\beta\sigma}-V-\Sigma_{\beta\sigma}^{0}\right)^{-1}\nonumber \\
 &  & \times\left(h\mathbb{G}_{\beta\alpha\alpha}^{\sigma\sigma\sigma}-\left\langle d_{\alpha\sigma}^{\dagger}d_{\beta\sigma}\right\rangle \right),\nonumber \\
\mathbb{G}_{\beta\beta\alpha}^{\bar{\sigma}\sigma\sigma} & = & \left(\hbar\omega-\varepsilon_{\beta\sigma}-U-\Sigma_{\beta\sigma}^{0}\right)^{-1}\nonumber \\
 &  & \times\left(h\mathbb{G}_{\beta\alpha\alpha}^{\bar{\sigma}\sigma\sigma}+h\mathbb{G}_{\beta\alpha\beta\alpha}^{\bar{\sigma}\bar{\sigma}\sigma\sigma}-h\mathbb{G}_{\alpha\beta\beta\alpha}^{\bar{\sigma}\bar{\sigma}\sigma\sigma}\right),\nonumber \\
\mathbb{G}_{\alpha\alpha\alpha}^{\bar{\sigma}\sigma\sigma} & = & \left(\hbar\omega-\varepsilon_{\alpha\sigma}-U-\Sigma_{\alpha\sigma}^{0}\right)^{-1}\nonumber \\
 &  & \times\left(\left\langle n_{\alpha\bar{\sigma}}\right\rangle +h\mathbb{G}_{\alpha\beta\alpha}^{\bar{\sigma}\sigma\sigma}+h\mathbb{G}_{\alpha\beta\alpha\alpha}^{\bar{\sigma}\bar{\sigma}\sigma\sigma}-h\mathbb{G}_{\beta\alpha\alpha\alpha}^{\bar{\sigma}\bar{\sigma}\sigma\sigma}\right),\nonumber \\
\mathbb{G}_{\beta\alpha\alpha}^{\sigma\sigma\sigma} & = & \left(\hbar\omega-\varepsilon_{\alpha\sigma}-V-\Sigma_{\alpha\sigma}^{0}\right)^{-1}\nonumber \\
 &  & \times\left(\left\langle n_{\beta\sigma}\right\rangle +h\mathbb{G}_{\alpha\beta\alpha}^{\sigma\sigma\sigma}\right),\nonumber \\
\mathbb{G}_{\beta\alpha\alpha}^{\bar{\sigma}\sigma\sigma} & = & \left(\hbar\omega-\varepsilon_{\alpha\sigma}-V-\Sigma_{\alpha\sigma}^{0}\right)^{-1}\nonumber \\
 &  & \times\left(\left\langle n_{\beta\bar{\sigma}}\right\rangle +h\mathbb{G}_{\beta\beta\alpha}^{\bar{\sigma}\sigma\sigma}+h\mathbb{G}_{\beta\alpha\alpha\alpha}^{\bar{\sigma}\bar{\sigma}\sigma\sigma}-h\mathbb{G}_{\alpha\beta\alpha\alpha}^{\bar{\sigma}\bar{\sigma}\sigma\sigma}\right),\nonumber \\
\mathbb{G}_{\beta\alpha\beta\alpha}^{\bar{\sigma}\bar{\sigma}\sigma\sigma} & = & \left(\hbar\omega-\varepsilon_{\beta\sigma}-\varepsilon_{\alpha\bar{\sigma}}+\varepsilon_{\beta\bar{\sigma}}\right)^{-1}\nonumber \\
 &  & \times\left(h\mathbb{G}_{\beta\beta\alpha}^{\bar{\sigma}\sigma\sigma}-h\mathbb{G}_{\alpha\beta\alpha}^{\bar{\sigma}\sigma\sigma}+h\mathbb{G}_{\beta\alpha\alpha\alpha}^{\bar{\sigma}\bar{\sigma}\sigma\sigma}\right),\nonumber \\
\mathbb{G}_{\beta\alpha\alpha\alpha}^{\bar{\sigma}\bar{\sigma}\sigma\sigma} & = & \left(\hbar\omega-\varepsilon_{\alpha\sigma}-\varepsilon_{\alpha\bar{\sigma}}+\varepsilon_{\beta\bar{\sigma}}\right)^{-1}\nonumber \\
 &  & \times\left(\left\langle d_{\beta\bar{\sigma}}^{\dagger}d_{\alpha\bar{\sigma}}\right\rangle -h\mathbb{G}_{\alpha\alpha\alpha}^{\bar{\sigma}\sigma\sigma}+h\mathbb{G}_{\beta\alpha\alpha}^{\bar{\sigma}\sigma\sigma}+h\mathbb{G}_{\beta\alpha\beta\alpha}^{\bar{\sigma}\bar{\sigma}\sigma\sigma}\right),\nonumber \\
\mathbb{G}_{\alpha\beta\beta\alpha}^{\bar{\sigma}\bar{\sigma}\sigma\sigma} & = & \left(\hbar\omega-\varepsilon_{\beta\sigma}-\varepsilon_{\beta\bar{\sigma}}+\varepsilon_{\alpha\bar{\sigma}}\right)^{-1}\nonumber \\
 &  & \times\left(h\mathbb{G}_{\alpha\beta\alpha}^{\bar{\sigma}\sigma\sigma}-h\mathbb{G}_{\beta\beta\alpha}^{\bar{\sigma}\sigma\sigma}+h\mathbb{G}_{\alpha\beta\alpha\alpha}^{\bar{\sigma}\bar{\sigma}\sigma\sigma}\right),\nonumber \\
\mathbb{G}_{\alpha\beta\alpha\alpha}^{\bar{\sigma}\bar{\sigma}\sigma\sigma} & = & \left(\hbar\omega-\varepsilon_{\alpha\sigma}-\varepsilon_{\beta\bar{\sigma}}+\varepsilon_{\alpha\bar{\sigma}}\right)^{-1}\nonumber \\
 &  & \times\left(\left\langle d_{\alpha\bar{\sigma}}^{\dagger}d_{\beta\bar{\sigma}}\right\rangle -h\mathbb{G}_{\beta\alpha\alpha}^{\bar{\sigma}\sigma\sigma}+h\mathbb{G}_{\alpha\alpha\alpha}^{\bar{\sigma}\sigma\sigma}+h\mathbb{G}_{\alpha\beta\beta\alpha}^{\bar{\sigma}\bar{\sigma}\sigma\sigma}\right).\nonumber \\
\label{eq:2-p GF case3}
\end{eqnarray}
This case is not different from the previous two in the sense that
a self consistent treatment is required.

\subsubsection{Approximation 4\label{sub:Case4}}

Finally we follow the derivation of Ref.~\onlinecite{Joshi2000}.
Here, the following approximations are made: (a) simultaneous tunneling
of electrons of opposite spins are neglected, (b) GFs mixing leads
and system operators are decoupled so $\hat{\mathbb{F}}\left(t,t'\right)=-\frac{i}{\hbar}\left\langle T_{C}c_{k\sigma}\left(t\right)n_{\alpha\tau}\left(t\right)d_{\beta\sigma}^{\dagger}\left(t'\right)\right\rangle \approx-\frac{i}{\hbar}t_{k\gamma}^{\sigma}\int\mbox{d}t_{1}\hat{g}_{k}\left(t,t_{1}\right)\left\langle T_{C}d_{\gamma\sigma}\left(t_{1}\right)n_{\alpha\tau}\left(t_{1}\right)d_{\beta\sigma}^{\dagger}\left(t'\right)\right\rangle $
(see discussion in subsection~\ref{sub:Case1}), and (c) $3$-particle
GFs are introduced in a mean-field way, i.e. $3$-particle GFs of
the form $-\frac{i}{\hbar}\left\langle T_{C}n_{\gamma\sigma'}\left(t\right)n_{\delta\tau}\left(t\right)d_{\alpha\sigma}\left(t\right)d_{\beta\sigma}^{\dagger}\left(t'\right)\right\rangle $
are decoupled to $-\frac{i}{\hbar}\left\langle n_{\gamma\sigma'}\left(t\right)\right\rangle \left\langle T_{C}n_{\delta\tau}\left(t\right)d_{\alpha\sigma}\left(t\right)d_{\beta\sigma}^{\dagger}\left(t'\right)\right\rangle -\frac{i}{\hbar}\left\langle n_{\delta\sigma}\left(t\right)\right\rangle \left\langle T_{C}n_{\gamma\sigma'}\left(t\right)d_{\alpha\sigma}\left(t\right)d_{\beta\sigma}^{\dagger}\left(t'\right)\right\rangle .$
The resulting equations are given by:
\begin{eqnarray}
\mathbb{G}_{\beta\beta\alpha}^{\bar{\sigma}\sigma\sigma} & = & \left(\hbar\omega-\varepsilon_{\beta\sigma}-U-V\left(\left\langle n_{\alpha\sigma}\right\rangle +\left\langle n_{\alpha\bar{\sigma}}\right\rangle \right)-\Sigma_{\beta\sigma}^{0}\right)^{-1}\nonumber \\
 & \times & \left(h\mathbb{G}_{\beta\alpha\alpha}^{\bar{\sigma}\sigma\sigma}+\left\langle n_{\beta\bar{\sigma}}\right\rangle V\left(\mathbb{G}_{\alpha\beta\alpha}^{\sigma\sigma\sigma}+\mathbb{G}_{\alpha\beta\alpha}^{\bar{\sigma}\sigma\sigma}\right)\right),\nonumber \\
\mathbb{G}_{\beta\alpha\alpha}^{\bar{\sigma}\sigma\sigma} & = & \left(\hbar\omega-\varepsilon_{\alpha\sigma}-U\left\langle n_{\alpha\bar{\sigma}}\right\rangle -V\left(\left\langle n_{\beta\sigma}\right\rangle +1\right)-\Sigma_{\alpha\sigma}^{0}\right)^{-1}\nonumber \\
 & \times & \left(\left\langle n_{\beta\bar{\sigma}}\right\rangle +h\mathbb{G}_{\beta\beta\alpha}^{\bar{\sigma}\sigma\sigma}+\left\langle n_{\beta\bar{\sigma}}\right\rangle \left(U\mathbb{G}_{\alpha\alpha\alpha}^{\bar{\sigma}\sigma\sigma}+V\mathbb{G}_{\beta\alpha\alpha}^{\sigma\sigma\sigma}\right)\right),\nonumber \\
\mathbb{G}_{\beta\alpha\alpha}^{\sigma\sigma\sigma} & = & \left(\hbar\omega-\varepsilon_{\alpha\sigma}-U\left\langle n_{\alpha\bar{\sigma}}\right\rangle -V\left(\left\langle n_{\beta\bar{\sigma}}\right\rangle +1\right)-\Sigma_{\alpha\sigma}^{0}\right)^{-1}\nonumber \\
 & \times & \left(\left\langle n_{\beta\sigma}\right\rangle +h\mathbb{G}_{\alpha\beta\alpha}^{\sigma\sigma\sigma}+\left\langle n_{\beta\sigma}\right\rangle \left(U\mathbb{G}_{\alpha\alpha\alpha}^{\bar{\sigma}\sigma\sigma}+V\mathbb{G}_{\beta\alpha\alpha}^{\bar{\sigma}\sigma\sigma}\right)\right),\nonumber \\
\mathbb{G}_{\alpha\beta\alpha}^{\bar{\sigma}\sigma\sigma} & = & \left(\hbar\omega-\varepsilon_{\beta\sigma}-U\left\langle n_{\beta\bar{\sigma}}\right\rangle -V\left(\left\langle n_{\alpha\sigma}\right\rangle +1\right)-\Sigma_{\beta\sigma}^{0}\right)^{-1}\nonumber \\
 & \times & \left(h\mathbb{G}_{\alpha\alpha\alpha}^{\bar{\sigma}\sigma\sigma}+\left\langle n_{\alpha\bar{\sigma}}\right\rangle \left(U\mathbb{G}_{\beta\beta\alpha}^{\bar{\sigma}\sigma\sigma}+V\mathbb{G}_{\alpha\beta\alpha}^{\sigma\sigma\sigma}\right)\right),\nonumber \\
\mathbb{G}_{\alpha\alpha\alpha}^{\bar{\sigma}\sigma\sigma} & = & \left(\hbar\omega-\varepsilon_{\alpha\sigma}-U-V\left(\left\langle n_{\beta\bar{\sigma}}\right\rangle +\left\langle n_{\beta\sigma}\right\rangle \right)-\Sigma_{\alpha\sigma}^{0}\right)^{-1}\nonumber \\
 & \times & \left(\left\langle n_{\alpha\bar{\sigma}}\right\rangle +h\mathbb{G}_{\alpha\beta\alpha}^{\bar{\sigma}\sigma\sigma}+\left\langle n_{\alpha\bar{\sigma}}\right\rangle V\left(\mathbb{G}_{\beta\alpha\alpha}^{\sigma\sigma\sigma}+\mathbb{G}_{\beta\alpha\alpha}^{\bar{\sigma}\sigma\sigma}\right)\right),\nonumber \\
\mathbb{G}_{\alpha\beta\alpha}^{\sigma\sigma\sigma} & = & \left(\hbar\omega-\varepsilon_{\beta\sigma}-U\left\langle n_{\beta\bar{\sigma}}\right\rangle -V\left(\left\langle n_{\alpha\bar{\sigma}}\right\rangle +1\right)-\Sigma_{\beta\sigma}^{0}\right)^{-1}\nonumber \\
 & \times & \left(h\mathbb{G}_{\beta\alpha\alpha}^{\sigma\sigma\sigma}-\left\langle d_{\alpha\sigma}^{\dagger}d_{\beta,\sigma}\right\rangle +\left\langle n_{\alpha\sigma}\right\rangle \left(U\mathbb{G}_{\beta\beta\alpha}^{\bar{\sigma}\sigma\sigma}+V\mathbb{G}_{\alpha\beta\alpha}^{\bar{\sigma}\sigma\sigma}\right)\right).\nonumber \\
\label{eq:2-p GF case4}
\end{eqnarray}
As before, the different expectation values ($\left\langle n_{\gamma\tau}\right\rangle =\left\langle d_{\gamma\tau}^{\dagger}d_{\gamma\tau}\right\rangle $
and $\left\langle d_{\alpha\sigma}^{\dagger}d_{\beta\sigma}\right\rangle $)
are obtained by integrating the relevant lesser GF, which is calculated
from the contour ordered GF.

\section{Results and discussion\label{sec:Results}}

\begin{figure}[t]
\centering{}\includegraphics[width=8.5cm]{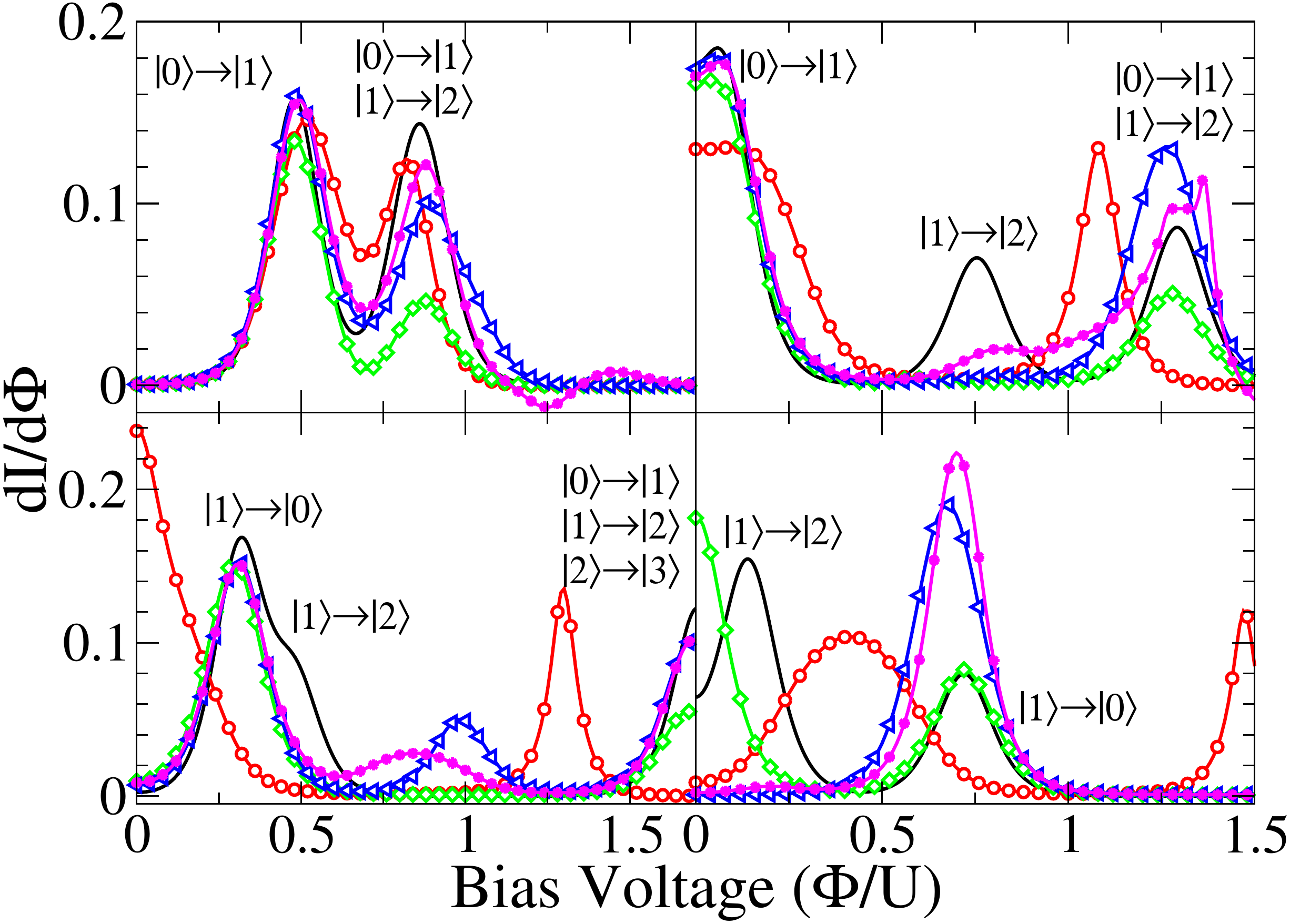}\caption{\label{fig:sym_bridge}(color online) Plots of the differential conductance
versus the bias voltage for the symmetric bridge ( $\varepsilon_{\alpha\uparrow}=\varepsilon_{\alpha\downarrow}=\varepsilon_{\beta\uparrow}=\varepsilon_{\beta\downarrow}=0.35U$)
for $V=0$. Upper left, upper right, lower left and lower right panels
correspond to $h=0.1U$, \uline{$0.3U$}, $0.5U$ and $0.7U$,
respectively. Black curves corresponds to results based on the ME.
Red (circles), green (diamonds), blue (triangles) and magenta (stars)
correspond to the results obtained by approximation schemes $1$ to
$4$, respectively. The notation $\left|i\right\rangle \rightarrow\left|j\right\rangle $
indicates that the conductance peak calculated by means of ME corresponds
to a transition form the $n_{i}$-particle states to any of the $n_{j}$-particle
states. The remaining model parameters were $\Gamma_{L\alpha}^{\uparrow}=\Gamma_{L\alpha}^{\downarrow}=\Gamma_{R\beta}^{\uparrow}=\Gamma_{R\beta}^{\downarrow}=0.015U$,
$\Gamma_{L\beta}^{\uparrow}=\Gamma_{L\beta}^{\downarrow}=\Gamma_{R\alpha}^{\uparrow}=\Gamma_{R\alpha}^{\downarrow}=0$,
and $\beta^{-1}=U/40$.}
\end{figure}

At this point we wish to examine the different approximations and
find which one leads to a system GF that describes the double QD more
accurately or at least qualitatively. As we are interested in transport
properties of a system weakly coupled to the macroscopic leads we
will compare the results obtained from the different approximations
to the ones calculated using the many-particle ME approach.~\citep{Oppenheim1977}
Under these assumptions, the ME is believed to be fairly accurate.~\citep{Beenakker1991,Datta1994}
We wish to note that the EOM approach is not limited to the weak coupling
case. As a measure of the quality of the approximations we chose to
calculate the differential conductance, $\mbox{d}I/\mbox{d}\Phi$.
In the many-particle picture, in the wide band limit (where the interaction
with the leads only broadens the energy levels of the system without
introducing any spectral shift), we expect to observe peaks in the
differential conductance at values of the bias voltage that correspond
to $\mu_{L/R}-E_{f}=\pm e\frac{\Phi}{2}\approx\Delta E\left(N\right)=E\left(N\right)-E\left(N-1\right)$,
where $E_{f}$ is the equilibrium Fermi energy of the electrodes (throughout
taken to be zero), $E\left(N\right)$ is the energy of the many-particle
state with $N$ electrons of the unperturbed system, and $\Phi$ is
the applied voltage. The voltage can be applied symmetrically to both
leads (i.e., $\mu_{L}=E_{f}+e\frac{\Phi}{2}$ and $\mu_{R}=E_{f}-e\frac{\Phi}{2}$),
or asymmetrically ($\mu_{L}=E_{f}+e\Phi$ and $\mu_{R}=E_{f}$). In
the present study we have used the symmetric version. 

The differential conductance is derived from differentiating the steady
state current with respect to the bias voltage and was evaluated from
the Meir-Wingreen formula:~\citep{Meir1992}
\begin{eqnarray}
I & = & \frac{ie}{2\pi\hbar}\int\mbox{d}\varepsilon\left(\mbox{Tr}\left\{ f_{L}\left(\varepsilon-\mu_{L}\right)\boldsymbol{\Gamma}_{L}\left(\varepsilon\right)\right.\right.\nonumber \\
 &  & \times\left.\left.\left(\mathbf{G}^{r}\left(\varepsilon\right)-\mathbf{G}^{a}\left(\varepsilon\right)\right)\right\} +\mbox{Tr}\left\{ \boldsymbol{\Gamma}_{L}\mathbf{G}^{<}\left(\varepsilon\right)\right\} \right).\label{eq:I1}
\end{eqnarray}
In the above, $\mathbf{G}^{r}\left(\varepsilon\right)$, $\mathbf{G}^{a}\left(\varepsilon\right)$
and $\mathbf{G}^{<}\left(\varepsilon\right)$ are the the retarded,
advanced and lesser GFs, respectively and $\boldsymbol{\Gamma}_{L}$
is the matrix coupling of the system to the left reservoir, with elements
$\left(\Gamma_{L}\right)_{\alpha\alpha}^{\sigma}=2\pi\underset{k\in L}{\sum}\delta\left(\varepsilon-\varepsilon_{k\sigma}\right)\left|t_{k\alpha}^{\sigma}\right|^{2}$.
The resulting EOMs were solved self-consistently in Fourier space
with a frequency discretization of $\mbox{d}\omega=0.0005U$ over
$32,768$ grid points. Depending on the approximation, $15-150$ self-consistent
iterations were required to converge the results. Convergence was
declared when the population values $\left(\left\langle n_{m\tau}\right\rangle \right)$
at subsequent iterations did not change within a predefined tolerance
value chosen to be $10^{-6}$. For each set of calculations symmetrization
routine was applied to restore the symmetry relations of the GFs.~\citep{Levy2013}

\subsection{Symmetric bridge\label{sub:Symmetric-bridge}}

\begin{figure}[t]
\centering{}\includegraphics[width=8.5cm]{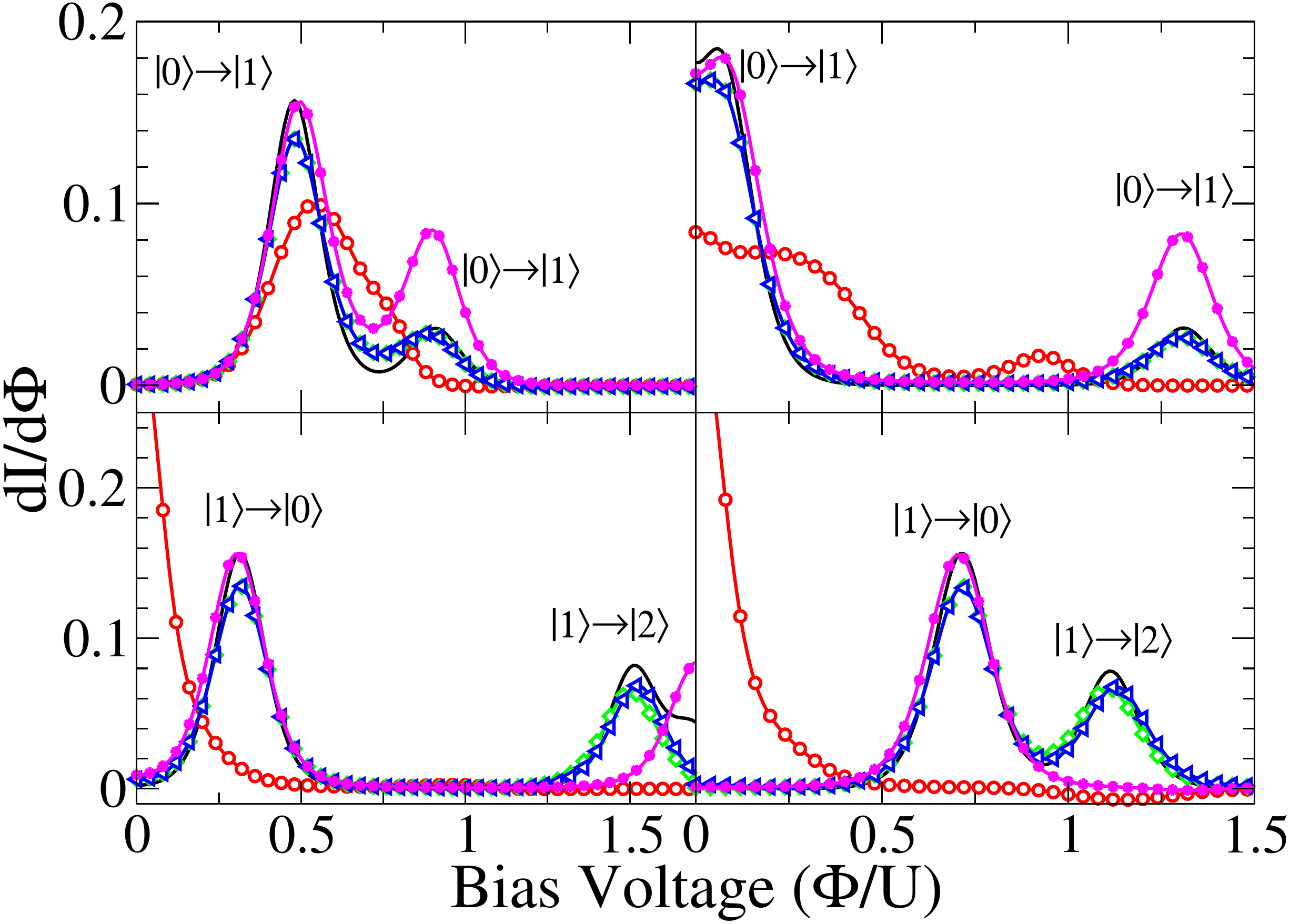}\caption{\label{fig:sym_bridge_V}(color online) Same as figure~\ref{fig:sym_bridge}
but for $V=0.8U$.}
\end{figure}

The transport through the double QD system can be classified into
symmetric and asymmetric bridge setups, with or without inter-dot
repulsion term, $V$. In this subsection we first consider the symmetric
setup in which $\varepsilon_{\alpha\uparrow}=\varepsilon_{\alpha\downarrow}=\varepsilon_{\beta\uparrow}=\varepsilon_{\beta\downarrow}=\varepsilon=0.35U$.
The remaining model parameters were taken to be $\Gamma_{L\alpha}^{\uparrow}=\Gamma_{L\alpha}^{\downarrow}=\Gamma_{R\beta}^{\uparrow}=\Gamma_{R\beta}^{\downarrow}=0.015U$,
$\Gamma_{L\beta}^{\uparrow}=\Gamma_{L\beta}^{\downarrow}=\Gamma_{R\alpha}^{\uparrow}=\Gamma_{R\alpha}^{\downarrow}=0$,
and $\beta^{-1}=U/40$. In figure~\ref{fig:sym_bridge} we plot the
differential conductance for a symmetric bridge for different values
of the hopping term $h$. We set the inter-dot repulsion $V=0$. The
black curves (solid line) are the results obtained by fully diagonalizing
the bare system $\hat{H}_{s}$ (and solving the ME). The other curves
represent the outcome of the NEGF formalism within the different closure
approximations. We also label the different peaks in the differential
conductance, obtained via the ME approach, with the corresponding
transitions between many-body states, i.e., $\left|0\right\rangle \rightarrow\left|1\right\rangle $
corresponds to transitions from an empty system to a system with a
single electron, etc. In the single particle GF formalism, peaks in
the differential conductance will occur at the poles of the calculated
GF. Hence a good approximation is one that will produce single particle
GF with poles at the position of the many-particle transitions.

For the smallest value of $h$, we find that all approximations agree
qualitatively with the ME approach. When the value of $h$ is increased
it is clear that approximation $1$ (red circles) breaks down, implying
that this simple closure is insufficient to describe strong hopping
between the quantum dots. Approximations $2$ and $3$ (green diamonds
and blue triangles, respectively) do agree with the ME, but ``miss''
certain conductance peaks (e.g. the peaks at $\nicefrac{\Phi}{U}\approx0.75$
in the upper right panel, $\nicefrac{\Phi}{U}\approx0.5$ in the lower
left panel and the peak at $\nicefrac{\Phi}{U}\approx0.2$ in the
lower right panel), all of which correspond to transitions involving
a $2$-electron occupancy. Approximation $4$ (magenta stars), which
includes $3$-particle GFs at a mean-field limit, performs slightly
better in this respect. 

\begin{table}[t]
\centering{}%
\begin{tabular}{|>{\raggedright}p{25mm}|>{\raggedright}p{35mm}|}
\hline 
GF poles & Energy differences\tabularnewline
\hline 
\hline 
~

$\varepsilon-\left|h\right|$

$\varepsilon+\left|h\right|$ & $\mathbf{\Delta}E\mathbf{\left(1\right)}:$

$\,$

$\varepsilon\pm\left|h\right|$\tabularnewline
\hline 
~

$\varepsilon+\frac{\left(U+V\right)}{2}-S_{1}$

$\varepsilon+\frac{\left(U+V\right)}{2}+S_{1}$

$\varepsilon+V-\left|h\right|$

$\varepsilon+V+\left|h\right|$ & $\mathbf{\Delta}E\mathbf{\left(2\right)}:$

$\,$

$\varepsilon+\left(U+V\right)/2-S_{2}\pm\left|h\right|$

$\varepsilon+\left(U+V\right)/2+S_{2}\pm\left|h\right|$

$\varepsilon+V\pm\left|h\right|$

$\varepsilon+U\pm\left|h\right|$\tabularnewline
\hline 
\end{tabular}\caption{\label{tab:poles and energy}Left column: Location of the poles of
the unperturbed system's GF as calculated using the $2^{\mbox{nd}}$
approximation. Right column: The differences in energy between many-particle
states that differ by one electron, such that $\Delta E\left(N\right)=E\left(N\right)-E\left(N-1\right)$.
Her$S_{1}=\frac{1}{2}\sqrt{\left(U-V\right)^{2}+4h^{2}}$, and $S_{2}=\frac{1}{2}\sqrt{\left(U-V\right)^{2}+16h^{2}}$}
\end{table}

In figure~\ref{fig:sym_bridge_V} we present results for the differential
conductance obtained for the symmetric bridge where the inter-dot
repulsion, $V$, is included. All other parameter are identical to
those of figure~\ref{fig:sym_bridge}. In this case, we find that
approximation $1$ is not suitable even for small values of the hopping
term $h$ while approximation $4$ appears to work for low values
of $h<\frac{1}{3}U$ (both upper panels), however, it fails to capture
peaks resulting from transitions through $2$-electron occupancy at
higher values of $h$, as depicted in the lower panels of figure~\ref{fig:sym_bridge_V}.
We note that for this set of parameters, such transitions involving
$2$ electrons are absent for $h<\frac{1}{3}U$ in the bias voltage
studied. Approximations $2$ and $3$ agree very well with the ME
results for all values of $h$, even at values of the bias voltage
that correspond to transfer through $2$-electron states, in contrast
to the case where $V=0$ in which they fail to capture conductance
peaks involving $2$ electrons.

The performance of the different approximations can be rationalized
in terms of the pole structure of the unperturbed system GF, which
can be compared to the exact many-body energy differences between
many-particle states that differ in one electron. While it is possible
to carry out this analysis for all closure approximations, it is often
a tedious task. Thus, in what follows we provide such analysis for
the case of approximation $2$ only. The poles of the GF, the many-particle
energy levels and the differences in energy are summarized in table~\ref{tab:poles and energy}
(see Appendix for more details regarding the derivation of the poles
of the GF). As can be seen from table~\ref{tab:poles and energy},
conductance peaks corresponding to transitions $\left|0\right\rangle \rightarrow\left|1\right\rangle $
and $\left|1\right\rangle \rightarrow\left|0\right\rangle $, that
is, peaks appearing at the values of $\Delta E(1)$, are captured
by approximation $2$ since the GF has poles at the correct locations.
Higher excitations involving $2$ or more electron occupancies are
not fully or systematically accounted for by approximation $2$ (or
any of the other closures described in this paper, for this matter).
In general we find that such higher transitions are not captured by
approximation $2$ when $V\ll U$. For $V=U$ we find that $\Delta E\left(2\right)$
has $4$ different values: $\varepsilon+V\pm\left|h\right|$ and $\varepsilon+V\pm2\left|h\right|$,
concurrently the GF has poles at $\varepsilon+V\pm\left|h\right|$,
thus, some of the transitions involving the $2$-electron states $\left|N=2\right\rangle $
(particularly those with $E\left(N=2\right)=\varepsilon+V\pm\left|h\right|$)
are described by the NEGF. 

Following this short analysis we can now better explain the results
of figure~\ref{fig:sym_bridge_V}. For a large value of the inter-dot
repulsion ($V=0.8U$, thus $V\sim U$), one expects that the calculated
GF will capture the higher order transitions in the relevant bias
window and agree with the ME results. If one considers the second
peak $\left(\nicefrac{\Phi}{U}\approx1.1\right)$ in the lower right
panel of figure~\ref{fig:sym_bridge_V}, it results from transmission
through a many-particle level with $\left|N=2\right\rangle $. For
the symmetric bridge, this peak corresponds to $\Delta E=\left(2\varepsilon+\frac{1}{2}\left(U+V\right)-S_{2}\right)-\left(\varepsilon-\left|h\right|\right)=\varepsilon+\frac{1}{2}\left(U+V\right)-S_{2}+\left|h\right|$,
where $S_{2}=\frac{1}{2}\sqrt{\left(U-V\right)^{2}+16h^{2}}$. Under
the assumption that $V\sim U$, one can approximate $\Delta E\approx\varepsilon+V-\left|h\right|$,
whereas the GF has one of its poles at $P_{G}=\varepsilon+\frac{1}{2}\left(U+V\right)-S_{1}$,
where $S_{1}=\frac{1}{2}\sqrt{\left(U-V\right)^{2}+4h^{2}}$, which,
for $V\sim U$ can be approximated by $P_{G}\approx\varepsilon+V-\left|h\right|$.
For the studied parameters (see figure~\ref{fig:symmetricBridgeSketch})
we find $P_{G}=0.54289$ and $\Delta E=0.54643$, and indeed the differential
conductance based on approximation $2$ show a peak at twice this
value$\nicefrac{\Phi}{U}\approx1.1$. While for the case where $V=0$
this transition is overlooked. It is easy to verify that a similar
argument holds for the second peak in the lower left panel of figure~\ref{fig:sym_bridge_V}
as well.

\subsection{Asymmetric bridge\label{sub:Asymmetric-bridge}}

\begin{figure}[t]
\centering{}\includegraphics[width=8.5cm]{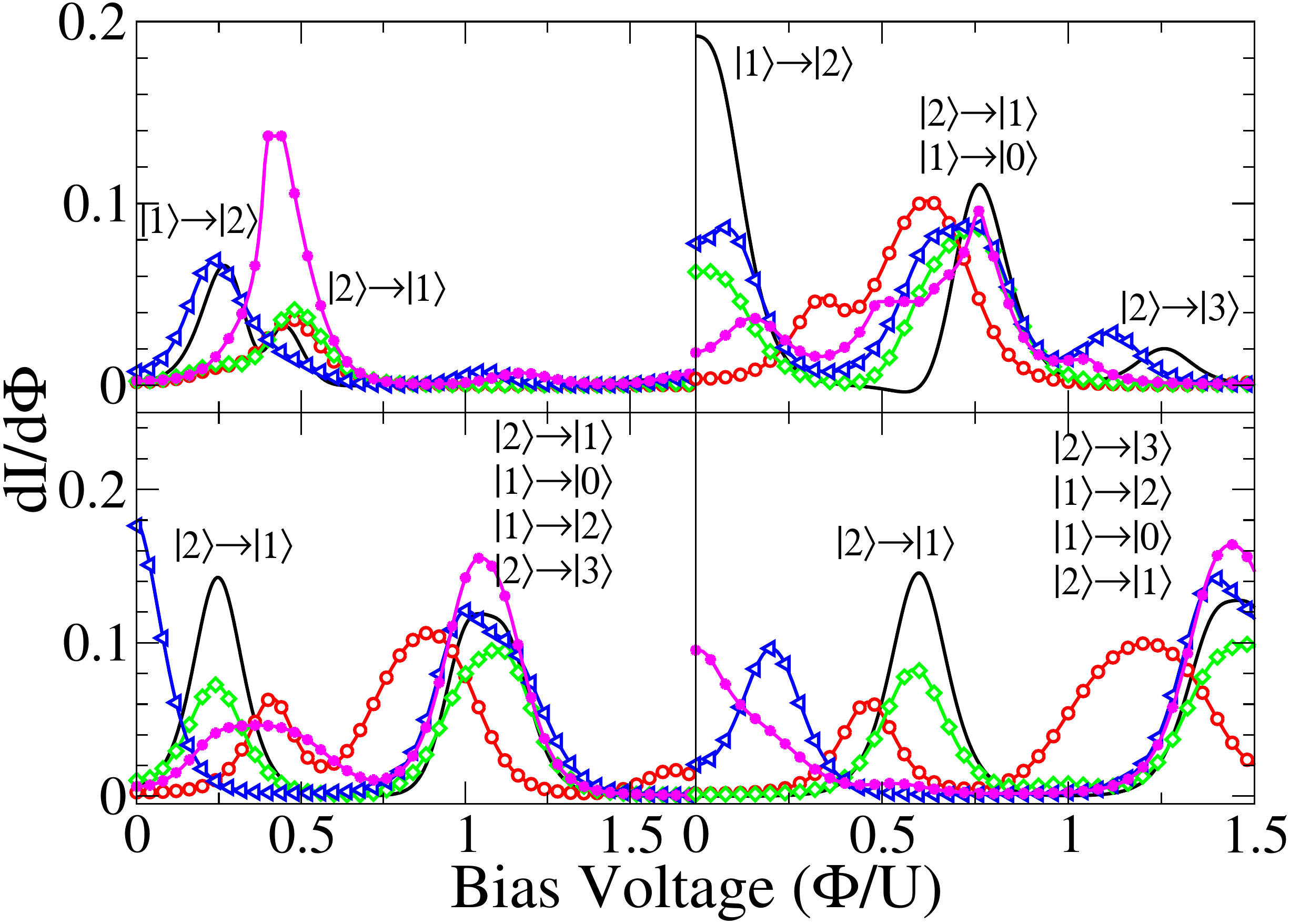}\caption{\label{fig:asymBridge}(color online) Plots of the differential conductance
versus the bias voltage for the asymmetric bridge ($\varepsilon_{\alpha\uparrow}=\varepsilon_{\alpha\downarrow}=0.15U$
and $\varepsilon_{\beta\uparrow}=\varepsilon_{\beta\downarrow}=-0.2U$)
for $V=0$. Upper left, upper right, lower left and lower right panels
correspond to $h=0.1U$, \uline{$0.3U$}, $0.5U$ and $0.7U$,
respectively. Black curves corresponds to results based on the ME.
Red (circles), green (diamonds), blue (triangles) and magenta (stars)
correspond to the results obtained by approximation schemes $1$ to
$4$, respectively. The notation $\left|i\right\rangle \rightarrow\left|j\right\rangle $
indicates that the conductance peak calculated by means of ME corresponds
to a transition form the $n_{i}$-particle states to any of the $n_{j}$-particle
states. The remaining model parameters were $\Gamma_{L\alpha}^{\uparrow}=\Gamma_{L\alpha}^{\downarrow}=\Gamma_{R\beta}^{\uparrow}=\Gamma_{R\beta}^{\downarrow}=0.015U$,
$\Gamma_{L\beta}^{\uparrow}=\Gamma_{L\beta}^{\downarrow}=\Gamma_{R\alpha}^{\uparrow}=\Gamma_{R\alpha}^{\downarrow}=0$,
and $\beta^{-1}=U/40$.}
\end{figure}

We now turn to discuss the case where $\varepsilon_{\alpha\sigma}\neq\varepsilon_{\beta\sigma}$
referred to as the asymmetric bridge. Once again we have calculated
the differential conductance using the $4$ different closure approximations
to the NEGF formalism and compared the results to the differential
conductance obtained by the ME. Analysis based on analytic expressions
for the poles of the GF or the many-particle energies of $\hat{H}_{S}$
is more difficult, and the expressions are not as compact as in the
symmetric case. The results for the poles of the GF within closure
approximation $2$ are given in the Appendix, while the many-body
energy differences were obtained numerically. 

\begin{figure}[t]
\centering{}\includegraphics[width=8.5cm]{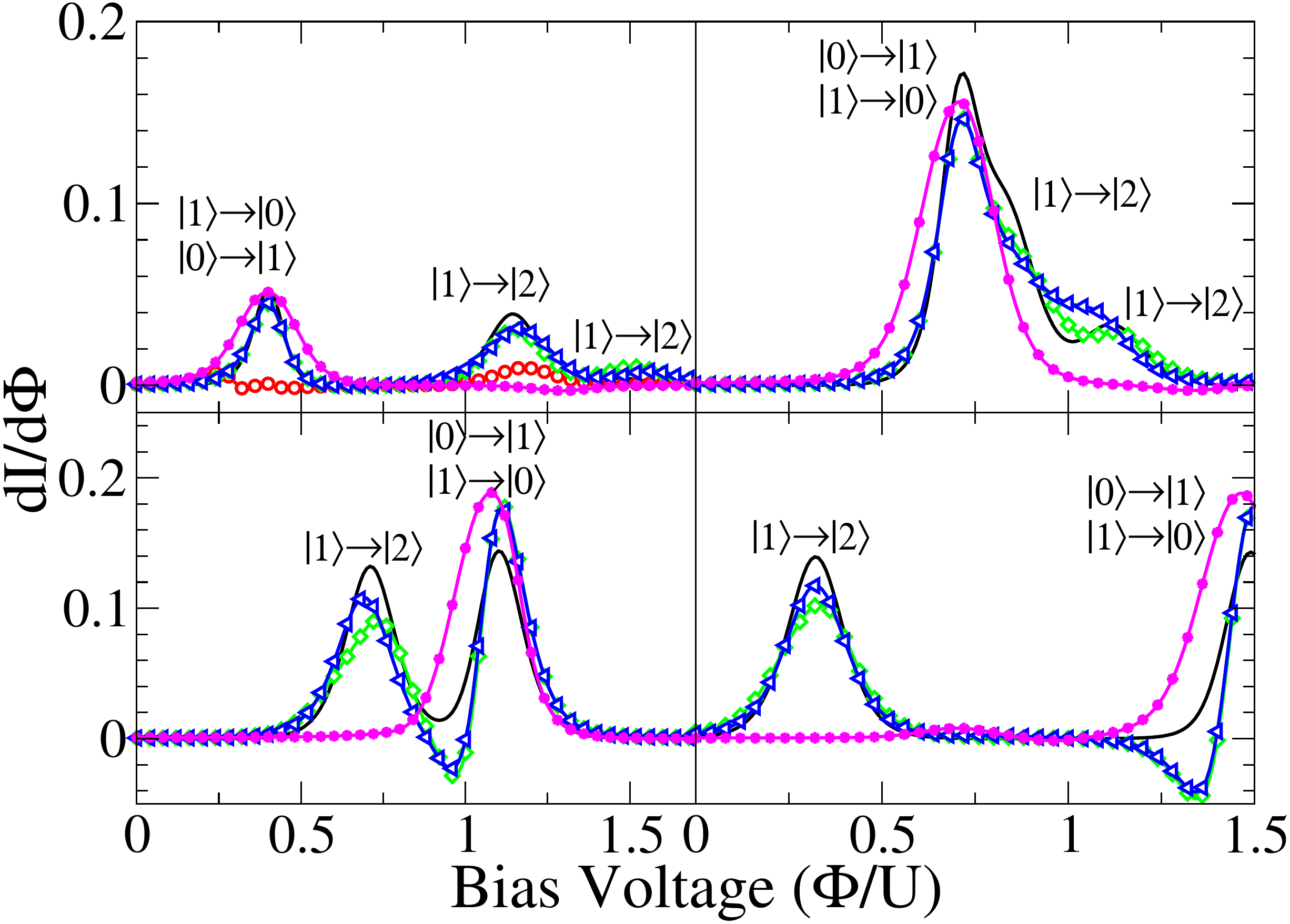}\caption{\label{fig:asymBridgeV}(color online) Same as figure~\ref{fig:asymBridge}
but for $V=0.8U$. Results obtained from approximation $1$ are only
presented for the case $h=0.1U$ (upper left panel) as we could not
converge it for higher values of $h$.}
\end{figure}

In figures~\ref{fig:asymBridge} and~\ref{fig:asymBridgeV} we plot
the differential conductance for the asymmetric bridge for different
values of the hopping term $h$ for $V=0$ and $V=0.8U$, respectively.
The on-site single particle energies were $\varepsilon_{\alpha\uparrow}=\varepsilon_{\alpha\downarrow}=0.15U$,
$\varepsilon_{\beta\uparrow}=\varepsilon_{\beta\downarrow}=-0.2U$
. The remaining model parameters are identical to those of the symmetric
bridge and were taken to be $\Gamma_{L\alpha}^{\uparrow}=\Gamma_{L\alpha}^{\downarrow}=\Gamma_{R\beta}^{\uparrow}=\Gamma_{R\beta}^{\downarrow}=0.015$U,
$\Gamma_{L\beta}^{\uparrow}=\Gamma_{L\beta}^{\downarrow}=\Gamma_{R\alpha}^{\uparrow}=\Gamma_{R\alpha}^{\downarrow}=0$,
and $\beta^{-1}=U/40$. As before, the black curves (solid line) corresponds
to the ME results. The other curves represent the outcome of the NEGF
formalism within the different closure approximations. We also label
the different peaks in the differential conductance with the corresponding
transitions between many-body states, i.e., $\left|0\right\rangle \rightarrow\left|1\right\rangle $
corresponds to transitions from an empty system to a system with a
single electron, etc. 

From figures~\ref{fig:asymBridge} and~\ref{fig:asymBridgeV} it
is obvious that approximations $1$ and $4$ do not perform as well
as approximations $2$ and $3$. We would like to note that approximations
$1$ and $4$ utilized a mean-field like approximation decoupling
the higher order GFs, while in approximations $2$ and $3$ higher
order GFs are ignored altogether. For all parameters studied in this
work (not all presented here), we find that approximation $2$ performed
better than all the other approximations, suggesting that including
higher order correlations in a mean field fashion or a more complete
treatment of the $2^{\mbox{nd}}$ order GFs is not advantageous. 

We find that approximations $2$ and $3$ predict negative differential
conductance at higher values of $h$, not obtained by the ME, as shown
in the lower panels of figure \ref{fig:asymBridgeV}. The dips occur
(in both cases) at values corresponding to the activation of the anti-bonding
single electron state. While this transition is suppressed in the
ME approach , it appears to be enhanced in the NEGF formalism.

\section{Concluding remarks\label{sec:Concluding-remarks} }

In this work we have assessed the validity of the EOM approach to
the NEGF formalism for an interacting system coupled to two macroscopic
leads. The interacting system consisted of two coupled quantum dots,
each with one electronic level (spin up and spin down), connected
serially, taking into account intra and inter-dot Coulomb interactions.
$4$ different closure approximations to the EOM, some are commonly
used in the literature and others have been developed here, were examined.
As a measure of the quality of the approximations we calculated the
differential conductance (derived by differentiating the steady state
current with respect to the bias voltage) and compared the results
to those obtained by the ME approach, which under the approximations
of weak coupling to the leads and high temperature provides accurate
results. Two different cases corresponding to a symmetric bridge $\left(\varepsilon_{\alpha\sigma}=\varepsilon_{\beta\sigma}\right)$
and an asymmetric bridge $\left(\varepsilon_{\alpha\sigma}\neq\varepsilon_{\beta\sigma}\right)$
with and without inter-dot Coulomb repulsion $\left(V\right)$, were
studied for different values of the inter-dot hopping term $h$. As
expected, we find that keeping more terms in the closure or including
higher order correlations in the EOM, does not necessarily improve
the approximation. As a rule of thumb, neglecting higher order GFs
(approximations $2$ and $3$) preforms better as compared to closures
that include such terms at a mean-field level (approximation $1$
and $4$).

To assess the performance of the different closure approximations,
we compared the pole structure of the uncoupled GF with the exact
results of the many-particle states. Focusing on approximation $2,$
which provides the overall best agreement in comparison to the ME
approach, we found that transitions involving only single electron
states were reproduced by the NEGF. However, when higher many-particle
states are involved, the accuracy of the approximation depends on
the strength of the Coulomb coupling $V$. In cases where $V\approx U$,
approximation $2$ also captures conductance peaks associated with
transitions involving two electron states. 

While all the approximations described in this work capture the Coulomb
blockade and the main characteristics of the steady state transport
for the single QD model (Anderson impurity model), they are not easily
expandable to systems with more complex nonequilibrium dynamics, such
as the double QD model (double Anderson model). In light of this,
the success of approximation $2$ for the double QD model does not
necessarily imply that it will provide quantitative results for a
more elaborate system. However, analysis of the poles of the resulting
bare GF and comparing them to the exact many-body energy differences
does provide a tool to assess the accuracy of a given approximate
closure and can be used for larger bridge systems even if these can
only be performed numerically.

\section{Acknowledgments\label{sec:Acknowledgments} }

This work was supported by the US-Israel Binational Science Foundation
and by the FP7 Marie Curie IOF project HJSC. TJL is grateful to the
Azrieli Foundation for the award of an Azrieli Fellowship.

\section*{Appendix \label{sec:Appendix}}

Using the assumptions described in Sec. \ref{sub:Case2}, the EOMs
for the NEGF of the unperturbed system are (in Fourier space):
\begin{eqnarray}
G_{\alpha\alpha}^{\sigma\sigma} & = & \left(\hbar\omega-\varepsilon_{\alpha\sigma}\right)^{-1}\nonumber \\
 &  & \times\left(1+hG_{\beta\alpha}^{\sigma\sigma}+U\mathbb{G}_{\alpha\alpha\alpha}^{\bar{\sigma}\sigma\sigma}+V\mathbb{G}_{\beta\alpha\alpha}^{\sigma\sigma\sigma}+V\mathbb{G}_{\beta\alpha\alpha}^{\bar{\sigma}\sigma\sigma}\right),\nonumber \\
G_{\beta\alpha}^{\sigma\sigma} & = & \left(\hbar\omega-\varepsilon_{\beta\sigma}\right)^{-1}\nonumber \\
 &  & \times\left(hG_{\alpha\alpha}^{\sigma\sigma}+U\mathbb{G}_{\beta\beta\alpha}^{\bar{\sigma}\sigma\sigma}+V\mathbb{G}_{\alpha\beta\alpha}^{\sigma\sigma\sigma}+V\mathbb{G}_{\alpha\beta\alpha}^{\bar{\sigma}\sigma\sigma}\right),\nonumber \\
\label{eq:ready1}
\end{eqnarray}
\begin{eqnarray}
\left(\hbar\omega-\varepsilon_{\beta\sigma}-V\right)\mathbb{G}_{\alpha\beta\alpha}^{\bar{\sigma}\sigma\sigma} & = & h\mathbb{G}_{\alpha\alpha\alpha}^{\bar{\sigma}\sigma\sigma},\nonumber \\
\left(\hbar\omega-\varepsilon_{\beta\sigma}-V\right)\mathbb{G}_{\alpha\beta\alpha}^{\sigma\sigma\sigma} & = & h\mathbb{G}_{\beta\alpha\alpha}^{\sigma\sigma\sigma}-\left\langle d_{\alpha\sigma}^{\dagger}d_{\beta,\sigma}\right\rangle ,\nonumber \\
\left(\hbar\omega-\varepsilon_{\beta\sigma}-U\right)\mathbb{G}_{\beta\beta\alpha}^{\bar{\sigma}\sigma\sigma} & = & h\mathbb{G}_{\beta\alpha\alpha}^{\bar{\sigma}\sigma\sigma},\nonumber \\
\left(\hbar\omega-\varepsilon_{\alpha\sigma}-U\right)\mathbb{G}_{\alpha\alpha\alpha}^{\bar{\sigma}\sigma\sigma} & = & \left\langle n_{\alpha\bar{\sigma}}\right\rangle +h\mathbb{G}_{\alpha\beta\alpha}^{\bar{\sigma}\sigma\sigma},\nonumber \\
\left(\hbar\omega-\varepsilon_{\alpha\sigma}-V\right)\mathbb{G}_{\beta\alpha\alpha}^{\sigma\sigma\sigma} & = & \left\langle n_{\beta\sigma}\right\rangle +h\mathbb{G}_{\alpha\beta\alpha}^{\sigma\sigma\sigma},\nonumber \\
\left(\hbar\omega-\varepsilon_{\alpha\sigma}-V\right)\mathbb{G}_{\beta\alpha\alpha}^{\bar{\sigma}\sigma\sigma} & = & \left\langle n_{\beta\bar{\sigma}}\right\rangle +h\mathbb{G}_{\beta\beta\alpha}^{\bar{\sigma}\sigma\sigma}.
\end{eqnarray}
It is clear that the equations for the $2$-particle GF close among
themselves, so a simple substitution yields: 
\begin{eqnarray}
\mathbb{G}_{\alpha\beta\alpha}^{\bar{\sigma}\sigma\sigma} & = & \left(\hbar\omega-\varepsilon_{\beta\sigma}-V-\frac{h^{2}}{\left(\hbar\omega-\varepsilon_{\alpha\sigma}-U\right)}\right)^{-1}\nonumber \\
 &  & \times\frac{h\left\langle n_{\alpha\bar{\sigma}}\right\rangle }{\left(\hbar\omega-\varepsilon_{\alpha\sigma}-U\right)}\nonumber \\
\mathbb{G}_{\alpha\beta\alpha}^{\sigma\sigma\sigma} & = & \left(\hbar\omega-\varepsilon_{\beta\sigma}-V-\frac{h^{2}}{\left(\hbar\omega-\varepsilon_{\alpha\sigma}-V\right)}\right)^{-1}\nonumber \\
 &  & \times\left(\frac{h\left\langle n_{\beta\sigma}\right\rangle }{\left(\hbar\omega-\varepsilon_{\alpha\sigma}-V\right)}-\left\langle d_{\alpha\sigma}^{\dagger}d_{\beta,\sigma}\right\rangle \right),\nonumber \\
\mathbb{G}_{\beta\beta\alpha}^{\bar{\sigma}\sigma\sigma} & = & \left(\hbar\omega-\varepsilon_{\beta\sigma}-U-\frac{h^{2}}{\left(\hbar\omega-\varepsilon_{\alpha\sigma}-V\right)}\right)^{-1}\nonumber \\
 &  & \times\frac{h\left\langle n_{\beta\bar{\sigma}}\right\rangle }{\left(\hbar\omega-\varepsilon_{\alpha\sigma}-V\right)},\nonumber \\
\mathbb{G}_{\alpha\alpha\alpha}^{\bar{\sigma}\sigma\sigma} & = & \left(\hbar\omega-\varepsilon_{\alpha\sigma}-U-\frac{h^{2}}{\left(\hbar\omega-\varepsilon_{\beta\sigma}-V\right)}\right)^{-1}\nonumber \\
 &  & \times\left\langle n_{\alpha\bar{\sigma}}\right\rangle \nonumber \\
\mathbb{G}_{\beta\alpha\alpha}^{\sigma\sigma\sigma} & = & \left(\hbar\omega-\varepsilon_{\alpha\sigma}-V-\frac{h^{2}}{\left(\hbar\omega-\varepsilon_{\beta\sigma}-V\right)}\right)^{-1}\nonumber \\
 &  & \times\left(\left\langle n_{\beta\sigma}\right\rangle -\frac{h\left\langle d_{\alpha\sigma}^{\dagger}d_{\beta,\sigma}\right\rangle }{\left(\hbar\omega-\varepsilon_{\beta\sigma}-V\right)}\right),\nonumber \\
\mathbb{G}_{\beta\alpha\alpha}^{\bar{\sigma}\sigma\sigma} & = & \left(\hbar\omega-\varepsilon_{\alpha\sigma}-V-\frac{h^{2}}{\left(\hbar\omega-\varepsilon_{\beta\sigma}-U\right)}\right)^{-1}\nonumber \\
 &  & \times\left\langle n_{\beta\bar{\sigma}}\right\rangle .
\end{eqnarray}
Define:
\begin{eqnarray}
x_{\alpha/\beta} & = & \varepsilon_{\alpha\sigma/\beta\sigma},\nonumber \\
x_{\alpha v/\beta v} & = & \varepsilon_{\alpha\sigma/\beta\sigma}+V,\nonumber \\
x_{\alpha u/\beta u} & = & \varepsilon_{\alpha\sigma/\beta\sigma}+U,
\end{eqnarray}
and rewrite the EOMs
\begin{eqnarray}
G_{\alpha\alpha}^{\sigma\sigma} & = & \left(\hbar\omega-x_{\alpha}\right)^{-1}\nonumber \\
 &  & \times\left(1+hG_{\beta\alpha}^{\sigma\sigma}+U\mathbb{G}_{\alpha\alpha\alpha}^{\bar{\sigma}\sigma\sigma}+V\mathbb{G}_{\beta\alpha\alpha}^{\sigma\sigma\sigma}+V\mathbb{G}_{\beta\alpha\alpha}^{\bar{\sigma}\sigma\sigma}\right),\nonumber \\
G_{\beta\alpha}^{\sigma\sigma} & = & \left(\hbar\omega-x_{\beta}\right)^{-1}\nonumber \\
 &  & \times\left(hG_{\alpha\alpha}^{\sigma\sigma}+U\mathbb{G}_{\beta\beta\alpha}^{\bar{\sigma}\sigma\sigma}+V\mathbb{G}_{\alpha\beta\alpha}^{\sigma\sigma\sigma}+V\mathbb{G}_{\alpha\beta\alpha}^{\bar{\sigma}\sigma\sigma}\right),\nonumber \\
\label{eq:ready1-1}
\end{eqnarray}
\begin{eqnarray}
\mathbb{G}_{\alpha\beta\alpha}^{\bar{\sigma}\sigma\sigma} & = & \frac{h\left\langle n_{\alpha\bar{\sigma}}\right\rangle }{\left(\left(\hbar\omega-x_{\beta v}\right)\left(\hbar\omega-x_{\alpha u}\right)-h^{2}\right)}\nonumber \\
\mathbb{G}_{\alpha\beta\alpha}^{\sigma\sigma\sigma} & = & \frac{h\left\langle n_{\beta\sigma}\right\rangle -\left\langle d_{\alpha\sigma}^{\dagger}d_{\beta,\sigma}\right\rangle \left(\hbar\omega-x_{\alpha v}\right)}{\left(\left(\hbar\omega-x_{\beta v}\right)\left(\hbar\omega-x_{\alpha v}\right)-h^{2}\right)},\nonumber \\
\mathbb{G}_{\beta\beta\alpha}^{\bar{\sigma}\sigma\sigma} & = & \frac{h\left\langle n_{\beta\bar{\sigma}}\right\rangle }{\left(\left(\hbar\omega-x_{\beta u}\right)\left(\hbar\omega-x_{\alpha v}\right)-h^{2}\right)},\nonumber \\
\mathbb{G}_{\alpha\alpha\alpha}^{\bar{\sigma}\sigma\sigma} & = & \frac{\left\langle n_{\alpha\bar{\sigma}}\right\rangle \left(\hbar\omega-x_{\beta v}\right)}{\left(\left(\hbar\omega-x_{\alpha u}\right)\left(\hbar\omega-x_{\beta v}\right)-h^{2}\right)}\nonumber \\
\mathbb{G}_{\beta\alpha\alpha}^{\sigma\sigma\sigma} & = & \frac{\left\langle n_{\beta\sigma}\right\rangle \left(\hbar\omega-x_{\beta v}\right)-h\left\langle d_{\alpha\sigma}^{\dagger}d_{\beta,\sigma}\right\rangle }{\left(\left(\hbar\omega-x_{\alpha v}\right)\left(\hbar\omega-x_{\beta v}\right)-h^{2}\right)},\nonumber \\
\mathbb{G}_{\beta\alpha\alpha}^{\bar{\sigma}\sigma\sigma} & = & \frac{\left\langle n_{\beta\bar{\sigma}}\right\rangle \left(\hbar\omega-x_{\beta u}\right)}{\left(\left(\hbar\omega-x_{\alpha v}\right)\left(\hbar\omega-x_{\beta u}\right)-h^{2}\right)}.\label{eq:ready2}
\end{eqnarray}
We now substitute the set of equations (\ref{eq:ready2}) into equations
(\ref{eq:ready1-1})
\begin{eqnarray}
G_{\alpha\alpha}^{\sigma\sigma} & = & \left(\left(\hbar\omega-x_{\alpha}\right)-\frac{h^{2}}{\left(\hbar\omega-x_{\beta}\right)}\right)^{-1}\nonumber \\
 &  & \times\left(1+\frac{hU\mathbb{G}_{\beta\beta\alpha}^{\bar{\sigma}\sigma\sigma}}{\left(\hbar\omega-x_{\beta}\right)}+\frac{hV\mathbb{G}_{\alpha\beta\alpha}^{\sigma\sigma\sigma}}{\left(\hbar\omega-x_{\beta}\right)}\right.\nonumber \\
 &  & +\frac{hV\mathbb{G}_{\alpha\beta\alpha}^{\bar{\sigma}\sigma\sigma}}{\left(\hbar\omega-x_{\beta}\right)}+U\mathbb{G}_{\alpha\alpha\alpha}^{\bar{\sigma}\sigma\sigma}\nonumber \\
 &  & \left.+V\mathbb{G}_{\beta\alpha\alpha}^{\sigma\sigma\sigma}+V\mathbb{G}_{\beta\alpha\alpha}^{\bar{\sigma}\sigma\sigma}\right),
\end{eqnarray}
Finally
\begin{eqnarray}
G_{\alpha\alpha}^{\sigma\sigma} & = & \left(\left(\hbar\omega-x_{\alpha}\right)\left(\hbar\omega-x_{\beta}\right)-h^{2}\right)^{-1}\nonumber \\
 &  & \times\left(1+\frac{h^{2}U\left\langle n_{\beta\bar{\sigma}}\right\rangle }{\left(\left(\hbar\omega-x_{\beta u}\right)\left(\hbar\omega-x_{\alpha v}\right)-h^{2}\right)}\right.\nonumber \\
 &  & +\frac{h^{2}V\left\langle n_{\beta\sigma}\right\rangle }{\left(\left(\hbar\omega-x_{\beta v}\right)\left(\hbar\omega-x_{\alpha v}\right)-h^{2}\right)}\nonumber \\
 &  & -\frac{hV\left\langle d_{\alpha\sigma}^{\dagger}d_{\beta,\sigma}\right\rangle \left(\hbar\omega-x_{\alpha v}\right)}{\left(\left(\hbar\omega-x_{\beta v}\right)\left(\hbar\omega-x_{\alpha v}\right)-h^{2}\right)}\nonumber \\
 &  & +\frac{h^{2}V\left\langle n_{\alpha\bar{\sigma}}\right\rangle }{\left(\left(\hbar\omega-x_{\beta v}\right)\left(\hbar\omega-x_{\alpha u}\right)-h^{2}\right)}\nonumber \\
 &  & +\frac{U\left\langle n_{\alpha\bar{\sigma}}\right\rangle \left(\hbar\omega-x_{\beta v}\right)\left(\hbar\omega-x_{\beta}\right)}{\left(\left(\hbar\omega-x_{\alpha u}\right)\left(\hbar\omega-x_{\beta v}\right)-h^{2}\right)}\nonumber \\
 &  & +\frac{\left\langle n_{\beta\sigma}\right\rangle V\left(\hbar\omega-x_{\beta}\right)\left(\hbar\omega-x_{\beta v}\right)}{\left(\left(\hbar\omega-x_{\alpha v}\right)\left(\hbar\omega-x_{\beta v}\right)-h^{2}\right)}\nonumber \\
 &  & -\frac{hV\left\langle d_{\alpha\sigma}^{\dagger}d_{\beta,\sigma}\right\rangle \left(\hbar\omega-x_{\beta}\right)}{\left(\left(\hbar\omega-x_{\alpha v}\right)\left(\hbar\omega-x_{\beta v}\right)-h^{2}\right)}\nonumber \\
 &  & \left.+\frac{\left\langle n_{\beta\bar{\sigma}}\right\rangle V\left(\hbar\omega-x_{\beta}\right)\left(\hbar\omega-x_{\beta u}\right)}{\left(\left(\hbar\omega-x_{\alpha v}\right)\left(\hbar\omega-x_{\beta u}\right)-h^{2}\right)}\right).\nonumber \\
\end{eqnarray}
From the last equation we see that the NEGF has poles at
\begin{eqnarray}
\left(\hbar\omega-x_{\alpha}\right)\left(\hbar\omega-x_{\beta}\right)-h^{2} & = & 0,\nonumber \\
\left(\hbar\omega-x_{\alpha v}\right)\left(\hbar\omega-x_{\beta v}\right)-h^{2} & = & 0,\nonumber \\
\left(\hbar\omega-x_{\alpha u}\right)\left(\hbar\omega-x_{\beta v}\right)-h^{2} & = & 0,\nonumber \\
\left(\hbar\omega-x_{\alpha v}\right)\left(\hbar\omega-x_{\beta u}\right)-h^{2} & = & 0,\nonumber \\
\end{eqnarray}
or equivalently 
\begin{eqnarray}
P_{G}^{1,2} & = & \frac{1}{2}\left[\left(x_{\alpha}+x_{\beta}\right)\pm\sqrt{\left(x_{\alpha}-x_{\beta}\right)^{2}+4h^{2}}\right],\nonumber \\
P_{G}^{3,4} & = & \frac{1}{2}\left[\left(x_{\alpha v}+x_{\beta v}\right)\pm\sqrt{\left(x_{\alpha v}-x_{\beta v}\right)^{2}+4h^{2}}\right],\nonumber \\
P_{G}^{5,6} & = & \frac{1}{2}\left[\left(x_{\alpha v}+x_{\beta u}\right)\pm\sqrt{\left(x_{\alpha v}-x_{\beta u}\right)^{2}+4h^{2}}\right],\nonumber \\
P_{G}^{7,8} & = & \frac{1}{2}\left[\left(x_{\alpha u}+x_{\beta v}\right)\pm\sqrt{\left(x_{\alpha u}-x_{\beta v}\right)^{2}+4h^{2}}\right].\nonumber \\
\end{eqnarray}

\bibliographystyle{apsrev4-1}
\bibliography{dqd}

\end{document}